\documentclass[aps,prd,superscriptaddress,twoside,showpacs,nofootinbib,10pt,floatfix,twocolumn,showpacs]{revtex4-1}%
\usepackage{verbatim}
\usepackage{bm}
\usepackage{tikz}
\usepackage{multirow}
\usepackage{amsmath}
\usepackage{gensymb}
\usepackage{graphicx}
\usepackage{epsfig}
\usepackage{braket}
\usepackage{subfigure}

\allowdisplaybreaks

\begin{document}
\title{Investigating full-heavy tetraquarks composed of $cc\bar{c}\bar{b}$ and $bb\bar{b}\bar{c}$ }
\author{Jing Hu}
\affiliation{
Department of Physics, Nanjing Normal University, Nanjing 210023, PR China
}
\author{Bing-Ran He}
\email[E-mail: ]{hebingran@njnu.edu.cn (Corresponding author)}
\affiliation{
Department of Physics, Nanjing Normal University, Nanjing 210023, PR China
}
\author{Jia-Lun Ping}
\email[E-mail: ]{jlping@njnu.edu.cn (Corresponding author)}
\affiliation{
Department of Physics, Nanjing Normal University, Nanjing 210023, PR China
}
\date{\today}
\begin{abstract}
The full-heavy tetraquarks $cc\bar{c}\bar{b}$ and $bb\bar{b}\bar{c}$ are systematically investigated within the chiral quark model. The meson-meson structure, diquark-antidiquark structure and K-structure are considered in this work. There is no bound state for $cc\bar{c}\bar{b}$ and $bb\bar{b}\bar{c}$ systems in $ IJ^{P}=00^{+},01^{+}$ and $ 02^{+}$ channels. However, for $cc\bar{c}\bar{b}$ system, three possible resonance states with energy of $ 10079~{\rm MeV} $, $ 10081~{\rm MeV} $ and $ 10177~{\rm MeV} $ are found in $ IJ^{P}=00^{+},01^{+}$ and $ 02^{+}$, respectively, and their decay width $ \Gamma$ are $6.7-8.4~{\rm MeV} $, $1.4-7.2~{\rm MeV} $ and $9.1-11.1~{\rm MeV} $. 
For $bb\bar{b}\bar{c}$ system, there also exist three possible resonance states with energy of $ 16474~{\rm MeV} $, $ 16474~{\rm MeV} $ and $ 16541~{\rm MeV} $ in $ IJ^{P}=00^{+},01^{+}$ and $ 02^{+}$, respectively, and the decay widths $ \Gamma$ of them are $2.2-6.1~{\rm MeV} $, $2.2-6.9~{\rm MeV} $ and $5.3-8.5~{\rm MeV} $.
$bc\bar{c}\bar{c}$ and $cb\bar{b}\bar{b}$ systems will have the same results as $cc\bar{c}\bar{b}$ and $bb\bar{b}\bar{c}$, respectively.
These full-heavy resonance states are worthy to be searched in the future experiments.
\end{abstract}
\maketitle

%##########
\section{\label{sec:level1}Introduction}
In the past few decades, many discoveries about exotic states inspired extensive interest in probing the structures of the multiquark hadrons. Gell-Man and Zweig have been proposed and invented multiquark states in the early quark models~\cite{Zweig:1964jf,Gell-Mann:1964ewy}. In 2003, Belle collaboration announced the observation of exotic state $X(3872)$~\cite{Belle:2003nnu}, and other experimental groups also found this exotic state~\cite{CDF:2003cab,D0:2004zmu,BaBar:2004oro}, which make the tetraquark state became a heated research area. After that, plenty of exotic states were observed and investigated, for review articles, please see~\cite{Brambilla:2010cs,Faccini:2012pj,Esposito:2014rxa,Esposito:2016noz,Chen:2016spr,Chen:2016qju,Guo:2017jvc,Olsen:2017bmm,Liu:2019zoy,Brambilla:2019esw,Chen:2022asf}. In 2017, the CMS collaboration implemented a benchmark measurement of the $\Upsilon$(1S) pair production at $ \sqrt{s} = 8 $ TeV in $ pp $ collision~\cite{CMS:2016liw}. However, no evidence has been found in the $\Upsilon$(1S)$\mu^+\mu^- $ invariant mass spectrum by the LHCb collaboration~\cite{LHCb:2018uwm}. Therefore, the fully-bottom tetraquark needs more experiments to confirm its signal. For fully-charm tetraquark systems, $J/\Psi$ pair production and double $c\bar{c}$ production have also been observed in experiments~\cite{LHCb:2011kri,Belle:2002tfa}. The $J/\Psi J/\Psi$ and $\eta_{c}(1S)\eta_{c}(1S)$ channels are suggested to search for the doubly hidden-charm. In 2020, the LHCb collaboration reported their elementary conclusion on the observations of full-charm states, a narrower structure at 6.9 GeV with significance about 5$\sigma$, a broad structure in the range 6.2 to 6.8 GeV, and there is also a hint for a structure around 7.2 GeV~\cite{LHCb:2020bwg}. These experimental results greatly stimulated much interest in full-heavy tetraquark states.
In 2022, based on all the proton-proton collisions collected from 2016 to 2018, the CMS collaboration has observed three structures X(6600), X(6900), and X(7300) in the invariant mass spectra of the charm quarks~\cite{CMS:2022yhl}.

In 1975, Y. Iwasaki proposed that full-charm tetraquark with the mass of about 6.0 GeV or 6.2 GeV is a sharp resonance state~\cite{Iwasaki:1975pv}. R.J. Lloyd \textit{et al.} investigated $ cc\bar{c}\bar{c} $ states and find several close-lying bound states~\cite{Lloyd:2003yc}. Berezhnoy \textit{et al.} obtained the mass $ M_{0}(cc\bar{c}\bar{c})= 6124 $ MeV and $ M_{0}(bb\bar{b}\bar{b})= 18857 $ MeV without hyperfine splitting, which involving charmed and bottom tetraquarks, respectively~\cite{Berezhnoy:2011xn}. M. Karliner \textit{et al.} discovered $M(cc\bar{c}\bar{c})= 6192\pm25 $MeV and $ M(bb\bar{b}\bar{b})= 18826\pm25 $MeV for  $ J^{PC} = 0^{++} $ involving charmed and bottom tetraquarks, respectively~\cite{Karliner:2016zzc}. W. Chen \textit{et al.} investigated  $cc\bar{c}\bar{c}$ and $bb\bar{b}\bar{b}$ states by a moment QCD sum rule method, and concluded that the mass of $\Upsilon\Upsilon$ and $\eta_{b}\eta_{b}$ are below and close to the corresponding thresholds except one current of $ J^{PC} = 0^{++} $, while the mass of $cc\bar{c}\bar{c}$ all above the thresholds~\cite{Chen:2016jxd}. 
J.~Wu \textit{et al.} discussed the $QQ\bar{Q}\bar{Q}$ configuration, and the result shows that the lowest $ J^P=1^+ $ state of $cc\bar{c}\bar{b}$ system should be less stable than that of $bb\bar{b}\bar{c}$ as the $ cc $ interaction is stronger than the $ bb $ interaction~\cite{Wu:2016vtq}. M. N. Anwar \textit{et al.} found that the ground state $ bb\bar{b}\bar{b} $ tetraquark mass is predicted to be  $ M(bb\bar{b}\bar{b})= 18.72\pm0.02 $ GeV~\cite{Anwar:2017toa}. A. Esposito \textit{et al.} proposed a model based on the conjecture of a short range diquark repulsion in a compact tetraquark to estimate masses and widths of tetraquarks, and found that the $bb\bar{b}\bar{b}$ system with the mass of 18.8 GeV~\cite{Esposito:2018cwh}. 
G.J. Wang \textit{et al.} systematically calculate the mass spectra of the S-wave fully-heavy tetraquark states $cc\bar{c}\bar{c}$, $bb\bar{b}\bar{b}$ and $bb\bar{c}\bar{c}$ in two nonrelativistic quark models, and the numerical results shows that the ground $QQ\bar{Q}\bar{Q}$ tetraquark states are located above the corresponding scattering states~\cite{Wang:2019rdo}. M.S. Liu \textit{et al.} studied the all-heavy tetraquark systems, i.e., $cc\bar{c}\bar{c}$, $bb\bar{b}\bar{b}$, $bb\bar{c}\bar{c}$/$cc\bar{b}\bar{b}$, $bc\bar{c}\bar{c}$/$cc\bar{b}\bar{c}$, $bc\bar{b}\bar{b}$/$bb\bar{b}\bar{c}$, and $bc\bar{b}\bar{c}$ within a potential model by including the linear confining potential, Coulomb potential, and spin-spin interactions, and the results showed that all diquark-antidiquark states are found to have masses above the corresponding the thresholds of $(Q\bar{Q})-(Q\bar{Q})$ structure~\cite{Liu:2019zuc}. 
Moreover, there are also many researches about the narrow structure $X(6900)$ reported by LHCb in 2020~\cite{Wang:2017jtz,Yang:2020atz,Jin:2020jfc,Zhao:2020zjh,Yang:2021hrb,Liu:2021rtn,Wang:2021kfv,Lu:2020cns,Asadi:2021ids,Wu:2022qwd,Wan:2020fsk}, and more papers can be found in Ref.~\cite{Chen:2022asf} and reference there in. 
Some of these researches interprets $X(6900)$ as excited state of diquark-antidiquark structure in the $cc\bar{c}\bar{c}$ system by various quark model~\cite{Jin:2020jfc,Zhao:2020zjh,Yang:2021hrb,Liu:2021rtn,Wang:2021kfv,Lu:2020cns,Asadi:2021ids} and by QCD sum rules method \cite{Wu:2022qwd,Wan:2020fsk,Wang:2017jtz}, while other study suggest that the component of it contains the excited state of meson-meson structure mixing with diquark-antiquark structure~\cite{Yang:2020atz}.

In this paper, we investigate the possible resonance states of $cc\bar{c}\bar{b}$ and $bb\bar{b}\bar{c}$ systems in $IJ^{P}=00^{+}$, $01^{+}$, and $02^{+}$ channels by using a quark model. The four-body configurations, which refers to meson-meson structure, diquark-antidiquark structure, and K-structure, as well as their couplings, are considered in the calculation. In Section.~\ref{sec:level2}, we give the introduction of the construction of wave functions. The numerical results and discussions are shown in Section.~\ref{sec:level3}. The summary is given in Section.~\ref{sec:level4}.

%##########
\section{\label{sec:level2} The quark model and wave functions of $cc\bar{c}\bar{b}$ and $bb\bar{b}\bar{c}$ systems}
Quantum chromodynamics(QCD) is recognized as a fundamental theory of strong interaction and the basic theory of multiquark states. However, with non-perturbative properties of QCD in the low energy region, it is inappropriate to directly use QCD theory to solve specific problems, such as hadron-hadron interactions and multiquark states. Therefore, some researchers have proposed and developed some quark models to solve these problems in low-energy regions. 

For full-heavy tetraquark systems, the potential is composed of the confinement and the one-gluon-exchange. The gluonic potential
have various forms, such as linear confinement~\cite{Godfrey:1985xj,Yang:2020atz,Liu:2021rtn,Yang:2021hrb,Zhao:2020zjh,Wang:2021kfv,Asadi:2021ids}, square confinement~\cite{Deng:2017xlb,Jin:2020jfc}, and exponential confinement~\cite{Yang:2020atz,Lu:2020cns}. 
The mass spectrum of these models is all in consistent with experimental results, and they all explained $X(6900)$~\cite{Jin:2020jfc,Zhao:2020zjh,Yang:2020atz,Lu:2020cns,Yang:2021hrb,Liu:2021rtn,Wang:2021kfv,Asadi:2021ids}.  
In other words, the forms of the confinement and the one-gluon-exchange in full-heavy tetraquark systems need more experimental data to be determined. 
Here we use one of these forms~\cite{Yang:2020atz,Vijande:2004he} to study possible resonance state in $cc\bar{c}\bar{b}$ and $bb\bar{b}\bar{c}$ systems.

The Hamiltonian in this model for the present study is written as:
\begin{eqnarray}
H&=&\sum_{i=1}^4(m_i+\frac{p_i^2}{2m_i})-T_{CM}+\sum_{j>i=1}^4(V^{CON}_{ij}+V^{OGE}_{ij}),\,\,\,\,\,\,\,\,
\end{eqnarray}
where $m_{i}$ and $p_{i}$ refer to the mass and the kinetic of $ i$-{th} quark(antiquark), $T_{CM}$ is the kinetic energy of the center of mass in tetraquark system; $V_{ij}^{CON}$ and $V_{ij}^{OGE}$ are the interactions of the confinement and the one-gluon-exchange between the $i$-{th} and $j$-{th} quark, respectively. 
The $V_{ij}^{CON}$ is written as
\begin{eqnarray}\label{con}
V_{ij}^{CON}  &=& [ -a_{c}~({1-e^{-\mu_{c}r_{ij}}})+\Delta]\boldsymbol{\lambda}^c_{i}\cdot
\boldsymbol{\lambda}^c_{j}\,,
\end{eqnarray} 
and the $V_{ij}^{OGE}$ express as
\begin{eqnarray}\label{oge}
V^{OGE}_{ij}  &=&  \frac{\alpha_s}{4}{\boldsymbol\lambda}^{c}_i \cdot
{\boldsymbol\lambda}^{c}_j
[\frac{1}{r_{ij}}-\frac{1}{6m_{i}m_{j}}(\frac{e^{-\frac{r_{ij}}{r_{0}(\mu)}}}{r_{ij}r^{2}_{0}(\mu)}
\boldsymbol{\sigma}_i\cdot\boldsymbol{\sigma}_j)]\,,\;\;\\
\alpha_s&=&\frac{\alpha_0}{\ln(\frac{\mu^{2}+\mu^{2}_{0}}{\Lambda^{2}_{0}})},~~~ r_{0}(\mu)=\frac{\widehat{r}_{0}}{\mu}\,.
\end{eqnarray} 
where $ r_{ij} $ stands for the distance between the two quarks/antiquarks, and $\sigma$ indicates the SU(2) Pauli matrices, while $\lambda^{c}$ is SU(3) color Gell-Mann matrices, respectively. $\alpha_s$ is an effective scale-dependent running coupling. $ \mu $ is the reduced mass of two quarks, and it written as $ \mu_{ij}=\frac{m_{i}m_{j}}{m_{i}+m_{j}} $. The $a_{c}$, $\Delta$, $\mu_{0}$, $\Lambda_{0}$, $\mu_{c}$, and $\widehat{r}_{0}$ are parameters listed in Table~\ref{tab:table1}, and these parameters are taken from Ref.~\cite{Vijande:2004he}. The mass of mesons that related to the present work are shown in Table~\ref{tab:table2}.

\begin{table}
\caption{\label{tab:table1}Model parameters}
\begin{tabular}{cccc}
\hline \hline
&~~Quark masses~~&~~~$m_{c}$({\rm MeV})~~~&~~~~1752\\
&&~~~$m_{b}$({\rm MeV})~~~&~~~~5100\\
\hline
&~~Confinement~~&~~~$a_{c}$({\rm MeV})~~~&~~~~430\\
&&~~~$\Delta$({\rm MeV})~~~&~~~~181.1\\
&&~~~$\mu_{c}({\rm fm^{-1}})$~~~&~~~~0.70\\
%&&~~~$a_{s}$~~~&~~~~0.777\\
\hline
&~~OGE~~&~~~$\alpha_{0}$~~~&~~~~2.118\\
&&~~~$\Lambda_{0}(\rm fm^{-1})$~~~&~~~~0.113\\
&&~~~$\mu_{0}$({\rm MeV})~~~&~~~~36.976\\
&&~~~$\widehat{r}_{0}$({\rm MeV} $\rm fm$)~~~&~~~~28.170\\
\hline \hline
\end{tabular}
\end{table}

\begin{table}[ht]
\caption{\label{tab:table2}The masses of the mesons obtained from present quark model related to this work. Experimental values are taken from the Particle Data Group (PDG)~\cite{ParticleDataGroup:2022pth}.}
\begin{tabular}{ccc}
\hline \hline
~~~~~Meson~~~~~&$M_{the}\rm (MeV)$& ~~~~~$M_{exp}\rm (MeV)$~~~~~ \\ \hline
$\eta_{c}(1s)$       & 2989 & 2984 \\
$\eta_{c}(2s)$       & 3627 & 3638 \\
$J/\psi(1s)$   & 3097 & 3097 \\
$\psi(2s)$   & 3685 &  3686\\
$B_{c}(1s)$   & 6276 & 6274 \\
$B_{c}(2s)$   & 6857 & 6871 \\
$B^{*}_{c}(1s)$  & 6331 & -- \\
$B^{*}_{c}(2s)$  & 6887 & -- \\
$\eta_{b}(1s)$  & 9453 & 9399 \\
$\eta_{b}(2s)$  & 9985 & 9999 \\
$\Upsilon(1s)$  & 9505 & 9460\\
$\Upsilon(2s)$  & 10013 & 10023 \\\hline\hline
\end{tabular}
\end{table}

\begin{figure}[h]
\includegraphics[scale=1.0]{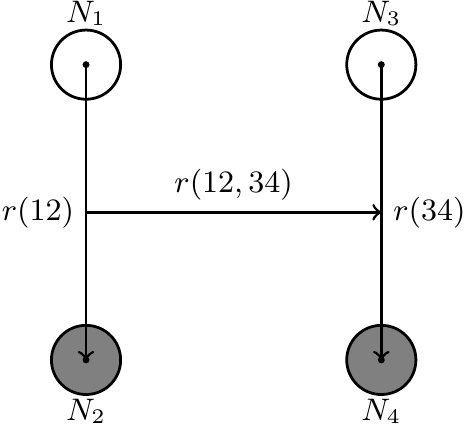}
\caption{\label{fig:1_H}Meson-meson structure.}
\end{figure}

\begin{figure}[h]
\includegraphics[scale=1.0]{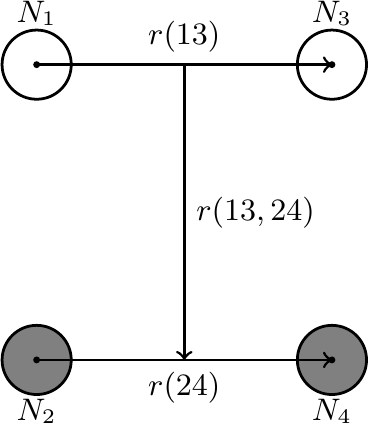}
\caption{\label{fig:2_di}Diquark-antidiquark structure.}
\end{figure}

\begin{figure}[h]
\includegraphics[scale=1.0]{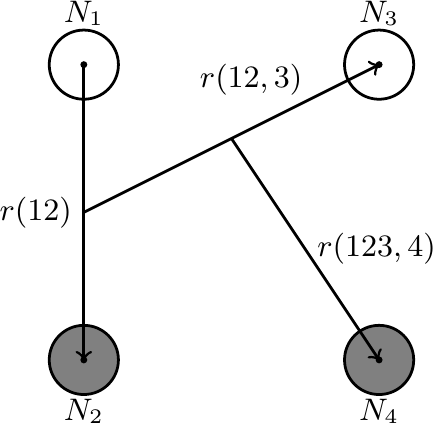}
\caption{\label{fig:3_K}K-structure.}
\end{figure}

According to different coupling methods, $cc\bar{c}\bar{b}$ and $bb\bar{b}\bar{c}$ systems have three structures are considered, i.e., meson-meson structure, diquark-antidiquark structure and K-structure, and are shown in Fig.~\ref{fig:1_H}, Fig.~\ref{fig:2_di} and Fig.~\ref{fig:3_K}. The hollow circles and black discs refer to quark and antiquark, respectively. The total wave functions of each structure are constructed by four parts: orbit, spin, flavor and color wave functions. The meson-meson structure and diquark-antidiquark structure are constructed by coupling two sub-clusters wave functions, while the K-structure is coupling a quark on the basis of quark-antiquark system and then coupling another antiquark. In the following discussions, the spin, flavor and color wave functions of meson-meson structure and K-structure are written in order of $N_{1}N_{2}N_{3}N_{4}$, while that of diquark-antidiquark structure is written in order of $N_{1}N_{3}N_{2}N_{4}$.

The total function of meson-meson structure ($ \Psi_{m} $), diquark-antidiquark structure ($ \Psi_{di} $) and K-structure ($ \Psi_{K} $) are written as,
\begin{eqnarray}
\Psi_{m} &=& {\cal A } \left[[\Phi^{L}_{m}\chi^{\sigma}_{m}]_{JM}\chi^{f}_{m}\chi^{c}_{m}\right] \nonumber\\
&=&{\cal A }\left[[[[\Phi_{l_{1}}(r_{12})\chi^{\sigma_{12}}][\Phi_{l_{2}}(r_{34})\chi^{\sigma_{34}}]]\Phi_{l_{3}}(r_{12,34})]^{JM}\right.\nonumber\\
&&\;\;\;\;\times\left.\chi^{f}_{m}\chi^{c}_{m}\right]\,,\label{eq:psi_m}\\
\Psi_{di} &=& {\cal A } \left[[\Phi^{L}_{di}\chi^{\sigma}_{di}]_{JM}\chi^{f}_{di}\chi^{c}_{di}\right] \nonumber\\
&=&{\cal A }\left[[[[\Phi_{l_{1}}(r_{13})\chi^{\sigma_{13}}][\Phi_{l_{2}}(r_{24})\chi^{\sigma_{24}}]]\Phi_{l_{3}}(r_{13,24})]^{JM}\right.\nonumber\\
&&\;\;\;\;\times\left.\chi^{f}_{di}\chi^{c}_{di}\right]\,,\label{eq:psi_di}\\
\Psi_{K}&=& {\cal A } [[\Phi^{L}_{K}\chi^{\sigma}_{K}]_{JM}\chi^{f}_{K}\chi^{c}_{K}] \nonumber\\
&=&{\cal A }\left[[[[\Phi_{l_{1}}(r_{12})\chi^{\sigma_{12}}]\Phi_{l_{2}}(r_{12,3})\chi^{\sigma_{12,3}}]\Phi_{l_{3}}(r_{123,4})]^{JM}\right.\nonumber\\
&&\;\;\;\;\times\left.\chi^{f}_{K}\chi^{c}_{K}\right]\,.\label{eq:psi_k}
\end{eqnarray}

For $cc\bar{c}\bar{b}$ and $bb\bar{b}\bar{c}$ systems, the two quarks with labels of 1 and 3 are identical particles, while the two anti-quarks with labels of 2 and 4 are not identical particles. Therefore, the antisymmetrization operator $ \cal A $ of meson-meson structure, diquark-antidiquark structure and K-structure is 
\begin{eqnarray}
	{\cal A }=1-(13)\,.
\end{eqnarray}
The $ \Phi^{L} $, $ \chi^{\sigma} $, $ \chi^{f} $ and $ \chi^{c} $ refers to orbit, spin, flavor and color wave function, respectively.

\subsection{Gaussian expansion method(GEM)}
This work focuses on radial excitation state of these tetraquark systems. The GEM has been successfully used in the calculation of few-body systems due to it just demands a small number of Gaussians to stabilize the system~\cite{Hiyama:2003cu}. Therefore, this work calculates the orbit wave function by GEM.  
In GEM, the radial part of orbit wave function of meson-meson structure, diquark-antidiquark structure and K-structure are all expanded by Gaussians: 
\begin{eqnarray}
\Phi_{n}^{r} &=& \sum_{n=1}^{n_{\max}}c_{n}\psi^{G}_{n}(r)\,,\\
\psi^{G}_{n}(r) &=&\frac{1}{\sqrt{4\pi}} N_{n}e^{-\nu_{n}r^{2}}\,,\\
N_{n} &=& \left[\dfrac{4(2\nu_{n})^{\frac{3}{2}}}{\sqrt{\pi}}\right]^{\frac{1}{2}}\,.
\end{eqnarray}
$N_{n}$ and $c_{n}$ means normalization constants and the variational parameters, respectively. $n_{\max}$ is the Gaussian number, while $r_{max}$ and $r_{min}$ are parameters of the Gaussian size $\nu_{n}$, and the formulas of $\nu_{n}$ are chosen according to the following geometric progression:
\begin{eqnarray}
a=\left(\dfrac{r_{max}}{r_{\min}}\right)^{\frac{1}{n_{\max}-1}},~~r_{n}=r_{\min}a^{n-1},~~\nu_{n}=\dfrac{1}{r_{n}^{2}}\,.\;\;
\end{eqnarray}
For $cc\bar{c}\bar{b}$ systems, the internal parameters clusters are chosen as $ r_{min}=0.01~(\rm fm),\ r_{max}=2~(\rm fm),\ n_{max}=12 $, and the parameters between the clusters are chosen as $ r_{min}=0.01~(\rm fm),\ r_{max}=6~(\rm fm),\ n_{max}=16 $ . 
For $bb\bar{b}\bar{c}$ systems, the internal parameters clusters are chosen as $ r_{min}=0.08~(\rm fm),\ r_{max}=2~(\rm fm),\ n_{max}=15 $, and the parameters between the clusters are chosen as $ r_{min}=0.01~(\rm fm),\ r_{max}=6~(\rm fm),\ n_{max}=16 $. 
To test the stability, a second calculation was performed, which takes $ r_{min}=0.01~(\rm fm),\ r_{max}=2~(\rm fm),\ n_{max}=15 $ for internal clusters and $ r_{min}=0.01~(\rm fm),\ r_{max}=6~(\rm fm),\ n_{max}=20 $ between the clusters. The result of the energy for ground states and excited states are the same as the first calculation. Hence, the energy calculated by the Gauss number used for the first time is enough to stabilize the present study.

\subsection{Spin wave function}
The spin wave functions of meson-meson structure, diquark-antidiquark structure and K-structure are denoted by the subscript $``\chi^{\sigma_{i}}_{S,S_{z}}"$, the $ S $ and $ S_{z} $ refers to spin angular momentum and its third component, and $ i $ means the number of spin wave functions. $ \alpha $ and $\beta $ represent spin wave functions of spin-up (1 0) and spin-down (0 1) wave functions. Tables~\ref{tab:table_meson} and \ref{tab:table_quark_baryon} lists the spin wave function of quark, meson and baryon, and Tables~\ref{tab:table8} and \ref{tab:table9} lists all spin wave function of meson-meson structure, diquark-antidiquark structure and K-structure.

\begin{table}[ht]
\caption{\label{tab:table_meson}The spin wave function of meson.}
\begin{tabular}{ccccccccccc}
\hline \hline
Meson&Spin wave function \\ \hline
$ \chi^{\sigma}_{1,1}(2) $ & $ \alpha\alpha $ \\ \hline
$ \chi^{\sigma}_{1,0}(2) $ & $ \dfrac{1}{\sqrt{2}}(\alpha\beta+\beta\alpha) $ \\ \hline
$ \chi^{\sigma}_{1,-1}(2) $ & $ \beta\beta $ \\ \hline
$ \chi^{\sigma}_{0,0}(2) $ & $ \dfrac{1}{\sqrt{2}}(\alpha\beta-\beta\alpha) $\\ 
\hline\hline
\end{tabular}
\end{table}

\begin{table}[ht]
\caption{\label{tab:table_quark_baryon}The spin wave function of quark and baryon.}
\scalebox{1.0}{
\begin{tabular}{ccccccccccc}
\hline \hline
Quark&Spin wave function \\ \hline		
$ \chi^{\sigma}_{\frac{1}{2},\frac{1}{2}}(1) $ & $ \alpha $ \\ \hline
$ \chi^{\sigma}_{\frac{1}{2},-\frac{1}{2}}(1) $ & $ \beta $ \\ \hline\hline
Baryon&Spin wave function \\ \hline
$ \chi^{\sigma}_{\frac{3}{2},\frac{3}{2}}(3) $ & $ \alpha\alpha\alpha $ \\ \hline
$ \chi^{\sigma}_{\frac{3}{2},\frac{1}{2}}(3) $ & $ \dfrac{1}{\sqrt{3}}(\alpha\alpha\beta+\alpha\beta\alpha+\beta\alpha\alpha) $ \\ \hline
$ \chi^{\sigma}_{\frac{3}{2},-\frac{1}{2}} (3) $ & $ \dfrac{1}{\sqrt{3}}(\alpha\beta\beta+\beta\alpha\beta+\beta\beta\alpha) $ \\ \hline
$ \chi^{\sigma}_{(\frac{1}{2})_{1},\frac{1}{2}} (3) $ & $ \dfrac{1}{\sqrt{2}}(\alpha\beta\alpha-\beta\alpha\alpha) $ \\ \hline
$ \chi^{\sigma}_{(\frac{1}{2})_{2},\frac{1}{2}}(3) $ & $ \dfrac{1}{\sqrt{6}}(2\alpha\alpha\beta-\alpha\beta\alpha-\beta\alpha\alpha) $ \\ \hline
$ \chi^{\sigma}_{(\frac{1}{2})_{1},-\frac{1}{2}} (3) $ & $ \dfrac{1}{\sqrt{2}}(\alpha\beta\beta-\beta\alpha\beta) $ \\ \hline
$ \chi^{\sigma}_{(\frac{1}{2})_{2},-\frac{1}{2}}(3) $ & $ \dfrac{1}{\sqrt{6}}(\alpha\beta\beta+\beta\alpha\beta-2\beta\beta\alpha) $ \\ \hline
$ \chi^{\sigma}_{\frac{3}{2},-\frac{3}{2}}(3)$ & $\beta\beta\beta $\\
\hline\hline
\end{tabular}}
\end{table}

\begin{table}[ht]
\caption{\label{tab:table8}The spin wave function of meson-meson structure and diquark-antidiquark structure.}
\scalebox{0.9}{
\begin{tabular}{ccccccccccc}
\hline \hline
Number&Spin wave function
\\ \hline
$ \chi^{\sigma_{1}}_{0,0}(4) $&$ \chi^{\sigma}_{0,0}(2)\chi^{\sigma}_{0,0}(2)$\\ \hline
$ \chi^{\sigma_{2}}_{0,0}(4) $ &$ \frac{1}{\sqrt{3}}[\chi^{\sigma}_{1,1}(2)\chi^{\sigma}_{1,-1}(2)-\chi^{\sigma}_{1,0}(2)\chi^{\sigma}_{1,0}(2)+\chi^{\sigma}_{1,-1}(2)\chi^{\sigma}_{1,1}(2)] $ \\ \hline
$ \chi^{\sigma_{3}}_{1,1}(4)$ & $\chi^{\sigma}_{0,0}(2)\chi^{\sigma}_{1,1}(2)$\\ \hline
$ \chi^{\sigma_{4}}_{1,1}(4) $ &$ \chi^{\sigma}_{1,1}(2)\chi^{\sigma}_{0,0}(2) $\\ \hline
$ \chi^{\sigma_{5}}_{1,1}(4) $ & $ \frac{1}{\sqrt{2}}[\chi^{\sigma}_{1,1}(2)\chi^{\sigma}_{1,0}(2)-\chi^{\sigma}_{1,0}(2)\chi^{\sigma}_{1,1}(2)] $\\ \hline
$ \chi^{\sigma_{6}}_{2,2}(4)  $& $ \chi^{\sigma}_{1,1}(2)\chi^{\sigma}_{1,1}(2) $\\ \hline\hline
\end{tabular}}
\end{table}

\begin{table}[ht]
\caption{\label{tab:table9}The spin wave function of K-structure.}
\scalebox{0.9}{
\begin{tabular}{ccccccccccc}
\hline \hline
Number&Spin wave function
\\ \hline
$ \chi^{\sigma_{7}}_{0,0}(4) $&$ \frac{1}{\sqrt{2}}[\chi^{\sigma}_{(\frac{1}{2})_{1},\frac{1}{2}} (3)\chi^{\sigma}_{\frac{1}{2},-\frac{1}{2}}(1)-\chi^{\sigma}_{(\frac{1}{2})_{1},-\frac{1}{2}}(3)\chi^{\sigma}_{\frac{1}{2},\frac{1}{2}}(1)]$\\ \hline
$ \chi^{\sigma_{8}}_{0,0}(4) $ &$ \frac{1}{\sqrt{2}}[\chi^{\sigma}_{(\frac{1}{2})_{2},\frac{1}{2}} (3)\chi^{\sigma}_{\frac{1}{2},-\frac{1}{2}}(1)-
\chi^{\sigma}_{(\frac{1}{2})_{2},-\frac{1}{2}} (3)\chi^{\sigma}_{\frac{1}{2},\frac{1}{2}}(1)] $ \\ \hline
$ \chi^{\sigma_{9}}_{1,1}(4)$ & $\chi^{\sigma}_{(\frac{1}{2})_{1},\frac{1}{2}} (3)\chi^{\sigma}_{\frac{1}{2},\frac{1}{2}}(1)$\\ \hline
$ \chi^{\sigma_{10}}_{1,1}(4) $ &$ \chi^{\sigma}_{(\frac{1}{2})_{2},\frac{1}{2}} (3)\chi^{\sigma}_{\frac{1}{2},\frac{1}{2}}(1)$\\ \hline
$ \chi^{\sigma_{11}}_{1,1}(4) $ & $ \sqrt{\frac{3}{4}}\chi^{\sigma}_{\frac{3}{2},\frac{3}{2}} (3)\chi^{\sigma}_{\frac{1}{2},-\frac{1}{2}}(1)-
\frac{1}{\sqrt{4}}\chi^{\sigma}_{\frac{3}{2},\frac{1}{2}} (3)\chi^{\sigma}_{\frac{1}{2},\frac{1}{2}}(1)$\\ \hline
$ \chi^{\sigma_{12}}_{2,2}(4)  $& $ \chi^{\sigma}_{\frac{3}{2},\frac{3}{2}} (3)\chi^{\sigma}_{\frac{1}{2},\frac{1}{2}}(1) $\\ \hline\hline
\end{tabular}}
\end{table}

\subsection{Flavor wave function}

The total flavor wave functions are
\begin{eqnarray}
\chi^{f_{1}}_{m}&=&Q\bar{Q}Q\bar{Q}\,,\\
\chi^{f_{2}}_{di}&=&QQ\bar{Q}\bar{Q}\,,
\end{eqnarray}
where $\chi^{f_{1}}_{m}$ and $\chi^{f_{2}}_{di}$ refers to flavor of meson-meson structure and diquark-antidiquark structure, while $\chi^{f_{3}}_{K}$ is same as $\chi^{f_{1}}_{m}$ and represents flavor of K-structure.

\subsection{Color wave function}
The color wave function of meson-meson structure, diquark-antidiquark structure and K-structure have six channels, and they were written in Table \ref{tab:table_color}.
\begin{table}[ht]
	\caption{\label{tab:table_color}The color wave function of meson-meson structure, diquark-antidiquark structure and K-structure.}
	\begin{tabular}{ccccccccccc}
		\hline \hline
		Number&Color wave function
		\\ \hline
$ \chi^{c_{1}}_{m} $&$\frac{1}{\sqrt{9}}(r\bar{r}r\bar{r}+r\bar{r}g\bar{g}+r\bar{r}b\bar{b}+g\bar{g}r\bar{r} +g\bar{g}g\bar{g}$\\
&$+g\bar{g}b\bar{b}+b\bar{b}r\bar{r}+b\bar{b}g\bar{g}+b\bar{b}b\bar{b})$\\
\hline
$ \chi^{c_{2}}_{m} $&$ \frac{1}{\sqrt{72}}(3r\bar{b}b\bar{r}+3r\bar{g}g\bar{r}+3g\bar{b}b\bar{g}+3b\bar{g}g\bar{b}+3g\bar{r}r\bar{g} $\\
&$+3b\bar{r}r\bar{b}+2r\bar{r}r\bar{r}+2g\bar{g}g\bar{g}+2b\bar{b}b\bar{b}-r\bar{r}g\bar{g} $\\
&$ -g\bar{g}r\bar{r}-b\bar{b}g\bar{g}-b\bar{b}r\bar{r}-g\bar{g}b\bar{b}-r\bar{r}b\bar{b}) $\\
\hline
$\chi^{c_{3}}_{di}$&$ \frac{1}{\sqrt{24}}(2rr\bar{r}\bar{r}+rg\bar{g}\bar{r}+rg\bar{r}\bar{g}+gr\bar{g}\bar{r}+rb\bar{b}\bar{r}  $\\
&$ +rb\bar{r}\bar{b}+br\bar{b}\bar{r}+br\bar{r}\bar{b}+gr\bar{r}\bar{g}+2gg\bar{g}\bar{g} $\\
&$ +gb\bar{b}\bar{g}+gb\bar{g}\bar{b}+bg\bar{b}\bar{g}+bg\bar{g}\bar{b}+2bb\bar{b}\bar{b}) $\\ \hline
$ \chi^{c_{4}}_{di} $&$\frac{1}{\sqrt{12}}(rg\bar{r}\bar{g}-rg\bar{g}\bar{r}+gr\bar{g}\bar{r}-gr\bar{r}\bar{g}+rb\bar{r}\bar{b}-rb\bar{b}\bar{r}  $\\
&$ +br\bar{b}\bar{r}-br\bar{r}\bar{b}+gb\bar{g}\bar{b}-gb\bar{b}\bar{g}+bg\bar{b}\bar{g}-bg\bar{g}\bar{b}) $
\\
\hline\hline
	\end{tabular}
\end{table}

The $\chi^{c_{1}}_{m}$ and $\chi^{c_{2}}_{m}$ refers to the color singlet-singlet ($1 \bigotimes 1$) and color octet-octet($8 \bigotimes 8$) configurations for meson-meson structure, while $\chi^{c_{3}}_{di}$ and $\chi^{c_{4}}_{di}$ represents the sextet-antisextet ($6 \bigotimes \bar{6}$) and triplet-antitriplet ($\bar{3} \bigotimes 3$) for diquark-antidiquark structure. What is more, the $\chi^{c_{5}}_{K}$ and $\chi^{c_{6}}_{K}$ are same as $\chi^{c_{1}}_{m}$ and $\chi^{c_{2}}_{m}$, which means two different color wave configurations for K-structure, and they are described as ``$ K_{1} $'' and ``$ K_{2} $'', respectively.

What is more, because $1 \bigotimes 1$ and $8 \bigotimes 8$ are orthonormal wave functions of four quark state. All other color wave functions could be expressed as a combination of $1 \bigotimes 1$ and $8 \bigotimes 8$. For example, the relation between $\bar{3} \bigotimes 3$, $6 \bigotimes \bar{6}$ and $1 \bigotimes 1$, $8 \bigotimes 8$ is:
\begin{eqnarray}\label{convert}
\begin{pmatrix}
	&\bar{3} \bigotimes 3&\\
	&6 \bigotimes \bar{6}&\\
\end{pmatrix}=
\begin{pmatrix}
\frac{1}{\sqrt{3}}&-\frac{\sqrt{2}}{\sqrt{3}}&\\
\frac{\sqrt{2}}{\sqrt{3}}&\frac{1}{\sqrt{3}}&\\
\end{pmatrix} 
\begin{pmatrix}
	&1 \bigotimes 1&\\
	&8 \bigotimes 8&\\
\end{pmatrix}\,.
\end{eqnarray}
Therefore, the $\bar{3} \bigotimes 3$ has $1/3$ of $1 \bigotimes 1$ and $2/3$ of $8 \bigotimes 8$, while $6 \bigotimes \bar{6}$ has $2/3$ of $1 \bigotimes 1$ and $1/3$ of $8 \bigotimes 8$. 

Classify all color wave functions to $1 \bigotimes 1$ and $8 \bigotimes 8$ have physical sense. We can separate color configurations $1 \bigotimes 1$ to infinitely far, because the expectation value of the operator $\boldsymbol{\lambda}^c_{i}\cdot\boldsymbol{\lambda}^c_{j}$ is zero, when $ i $ and $ j $ belongs to different cluster. 
However, if we try to separate color octet-octet $8 \bigotimes 8$ structure, the operator $\boldsymbol{\lambda}^c_{i}\cdot\boldsymbol{\lambda}^c_{j}$ provide attractive force, when $ i $ and $ j $ belongs to different cluster. As confinement potential existent in infinite far, the energy of $8 \bigotimes 8$ goes to an infinite value due to the color confinement. 
In this sense, the physical channel with the color configurations $\bar{3} \bigotimes 3$, $6 \bigotimes \bar{6}$ and $8 \bigotimes 8$ are ``compact state''.

Because $|{\psi_{1 \bigotimes 1}^{c}}\rangle\langle{\psi_{1 \bigotimes 1}^{c}}|+|{\psi_{8 \bigotimes 8}^{c}}\rangle\langle{\psi_{8 \bigotimes 8}^{c}}|=1$, here $\psi_{1 \bigotimes 1}^{c}$ and $\psi_{8 \bigotimes 8}^{c}$  corresponds to color $1 \bigotimes 1$ and $8 \bigotimes 8$ configurations, respectively. We insert it to each physical channel as:
\begin{eqnarray}
	\langle{\Psi_{i}}|{\Psi_{i}}\rangle=\langle{\Psi_{i}}|{\psi_{1 \bigotimes 1}^{c}}\rangle\langle{\psi_{1 \bigotimes 1}^{c}}|{\Psi_{i}}\rangle+\\\nonumber
	\langle{\Psi_{i}}|{\psi_{8 \bigotimes 8}^{c}}\rangle\langle{\psi_{8 \bigotimes 8}^{c}}|{\Psi_{i}}\rangle\,,
\end{eqnarray}
where $\Psi_{i}$ is a total wave function of a physical channel expressed in Eqs.~\eqref{eq:psi_m}-\eqref{eq:psi_k}.  So, the component of physical channel with $1 \bigotimes 1$ is calculated by:
\begin{eqnarray}\label{eq:color_1}
	P_{1 \bigotimes 1}=
	\dfrac{\sum_{i}\langle{\Psi_{i}}|{\psi_{1 \bigotimes 1}^{c}}\rangle\langle{\psi_{1 \bigotimes 1}^{c}}|{\Psi_{i}}\rangle}{\sum_{i}\langle{\Psi_{i}}|{\Psi_{i}}\rangle}\,,
\end{eqnarray}
and the component of physical channel with $8 \bigotimes 8$ is:
\begin{eqnarray}\label{eq:color_8}
	P_{8 \bigotimes 8}=
	\dfrac{\sum_{i}\langle{\Psi_{i}}|{\psi_{8 \bigotimes 8}^{c}}\rangle\langle{\psi_{8 \bigotimes 8}^{c}}|{\Psi_{i}}\rangle}{\sum_{i}\langle{\Psi_{i}}|{\Psi_{i}}\rangle}\,.
\end{eqnarray}
One can easily show that $P_{1 \bigotimes 1}+P_{8 \bigotimes 8}\equiv1$.

%##########
\section{\label{sec:level3}Results and discussions}
We investigate full-heavy tetraquarks $cc\bar{c}\bar{b}$ and $bb\bar{b}\bar{c}$ systems in three kind of quark structures, i.e., meson-meson structure, diquark-antidiquark structure and K-structure, and construct all possible physical channels according to
the quantum numbers $IJ^{P}=00^{+}$, $01^{+}$, and $02^{+}$. 
The physical channel and the mass are shown in Tables~\ref{tab:table3} and \ref{tab:table4}.  The first column of these tables shows the wave function of orbit, spin, flavor and color for each channel. We denote the total function by the subscript $[i,j,k]_{\rm type}$, $``i,j,k"$ indicates spin, flavor and color, and ``type" denotes meson-meson structure, diquark-antidiquark structure and K-structure, i.e., ``m'', ``di'' and ``K'', respectively. The second columns ``channels" enumerates the subscript $[\chi^{f_{j}}_{m}]_{\chi^{c_{k}}_{m}}^{S(N_{1},N_{2})\oplus S(N_{3},N_{4})}$ in meson-meson structure,  the subscript $[\chi^{f_{j}}_{di}]_{\chi^{c_{k}}_{di}}^{S(N_{1},N_{3})\oplus S(N_{2},N_{4})}$ in diquark-antidiquark structure, and the subscript $[\chi^{f_{j}}_{K}]_{\chi^{c_{k}}_{K}}^{S(N_{1},N_{2},N_{3})\oplus S(N_{4})}$ in K-structure, respectively. The columns headed with $E_{th}$ means the theoretical thresholds for two meson systems, which is the energy when separate two color singlet mesons infinitely far from each other. 

For a given $IJ^P$, the $E_{sc}$ refers to the energy of each single-channel, and the $E_{cc}$ denotes the lowest energies of the coupling of all channels, respectively~\cite{Yang:2021hrb}. 
Which are written as:
\begin{eqnarray}
	\bra{\Psi_{i}}H\ket{\Psi_{i}}\begin{pmatrix}c_i
	\end{pmatrix}=E_{sc}\langle{\Psi_{i}}|{\Psi_{i}}\rangle\begin{pmatrix}c_i
	\end{pmatrix}\,,
\end{eqnarray}
and, 
\begin{eqnarray}
	\begin{pmatrix}
		\langle{\Psi_{1}}|H|{\Psi_{1}}\rangle &\ldots& \langle{\Psi_{1}}|H|{\Psi_{n}}\rangle\\
		\ldots&\ldots&\ldots\nonumber\\
		\langle{\Psi_{n}}|H|{\Psi_{1}}\rangle &\ldots& \langle{\Psi_{n}}|H|{\Psi_{n}}\rangle
	\end{pmatrix}\begin{pmatrix}c_1\\
		\ldots\\
		c_n
	\end{pmatrix}=\\
	E_{cc}\begin{pmatrix}
		\langle{\Psi_{1}}|{\Psi_{1}}\rangle &\ldots& \langle{\Psi_{1}}|{\Psi_{n}}\rangle\\
		\ldots&\ldots&\ldots\\
		\langle{\Psi_{n}}|{\Psi_{1}}\rangle &\ldots& \langle{\Psi_{n}}|{\Psi_{n}}\rangle
	\end{pmatrix} \begin{pmatrix}c_1\\
		\ldots\\
		c_n
	\end{pmatrix}\,.
\end{eqnarray}
Where $ c_i $ is the eigenvector of corresponding energy.

\begin{table}[ht]
\caption{\label{tab:table3}The energies (in MeV) of all structures for tetraquarks $cc\bar{c}\bar{b}$.}
\begin{tabular}{ccccccccccc}
\hline \hline
&&~~~~~~~$IJ^{P}=00^{+}$&&\\ \hline
& $[i,j,k]_{\rm type}$ & Channel & $E_{th}$ & $E_{sc}$ & $E_{cc}$  \\ \hline
& $[1,1,1]_{m}$ & $[\eta_{c}B_{c}]^{0\oplus0}_{1 \bigotimes 1}$ & $9265$  & $9266$ & $9266 $ \\
& $[2,1,1]_{m}$ & $[J/\psi B^{*}_{c}]^{1\oplus1}_{1 \bigotimes 1}$ & $9427$ & $9428$ &  \\
& $[1,1,2]_{m}$ & $[\eta_{c}B_{c}]^{0\oplus0}_{8 \bigotimes 8}$ && $9673$ &  \\
& $[2,1,2]_{m}$ & $[J/\psi B^{*}_{c}]^{1\oplus1}_{8 \bigotimes 8}$ && $9663$ &  \\
& $[1,2,3]_{di}$ & $[cc\bar{c}\bar{b}]^{0\oplus0}_{6 \bigotimes \bar{6}}$ && $9660$ &  \\ 
& $[2,2,4]_{di}$ & $[cc\bar{c}\bar{b}]^{1\oplus1}_{\bar{3} \bigotimes 3}$ && $9685$ &  \\
& $[7,3,5]_{K}$ & $[c\bar{c}c\bar{b}]^{(\frac{1}{2})_{1}\oplus\frac{1}{2}}_{K_{1}}$ && $9572$ &$  $ &\\
& $[8,3,5]_{K}$ & $[c\bar{c}c\bar{b}]^{(\frac{1}{2})_{2}\oplus\frac{1}{2}}_{K_{1}}$ && $9662$ &\\
& $[7,3,6]_{K}$ & $[c\bar{c}c\bar{b}]^{(\frac{1}{2})_{1}\oplus\frac{1}{2}}_{K_{2}}$ && $9689$ &  \\
& $[8,3,6]_{K}$ & $[c\bar{c}c\bar{b}]^{(\frac{1}{2})_{2}\oplus\frac{1}{2}}_{K_{2}}$ && $9665$ &  \\ 
\hline \hline
&&~~~~~~~$IJ^{P}=01^{+}$&&\\ \hline
& $[i,j,k]_{\rm type}$ & Channel&$E_{th}$& $E_{sc}$ & $E_{cc}$  \\ \hline
& $[3,1,1]_{m}$ & $[\eta_{c}B^{*}_{c}]^{0\oplus1}_{1 \bigotimes 1}$ & $9320$ & $9320$ & $9320$ \\
& $[4,1,1]_{m}$ & $[J/\psi B_{c}]^{1\oplus0}_{1 \bigotimes 1}$ &$9373$ & $9373$ &  \\
& $[5,1,1]_{m}$ & $[J/\psi B^{*}_{c}]^{1\oplus1}_{1 \bigotimes 1}$ & $9427$ & $9428$ &  \\
& $[3,1,2]_{m}$ & $[\eta_{c}B^{*}_{c}]^{0\oplus1}_{8 \bigotimes 8}$ && $9668$ &  \\
& $[4,1,2]_{m}$ & $[J/\psi B_{c}]^{1\oplus0}_{8 \bigotimes 8}$ && $9667$ &  \\ 
& $[5,1,2]_{m}$ & $[J/\psi B^{*}_{c}]^{1\oplus1}_{8 \bigotimes 8}$ && $9660$ &  \\
& $[3,2,3]_{di}$ & $[cc\bar{c}\bar{b}]^{0\oplus1}_{6 \bigotimes \bar{6}}$ && $9658$ &  \\
& $[4,2,4]_{di}$ & $[cc\bar{c}\bar{b}]^{1\oplus0}_{\bar{3} \bigotimes 3}$ && $9683$ &  \\
& $[5,2,4]_{di}$ & $[cc\bar{c}\bar{b}]^{1\oplus1}_{\bar{3} \bigotimes 3}$ && $9691$ &  \\
& $[9,3,5]_{K}$ & $[c\bar{c}c\bar{b}]^{(\frac{1}{2})_{1}\oplus\frac{1}{2}}_{K_{1}}$ && $9578$ &$$&  \\
& $[10,3,5]_{K}$ & $[c\bar{c}c\bar{b}]^{(\frac{1}{2})_{2}\oplus\frac{1}{2}}_{K_{1}}$ && $9655$ &  \\ 
& $[11,3,5]_{K}$ & $[c\bar{c}c\bar{b}]^{\frac{3}{2}\oplus\frac{1}{2}}_{K_{1}}$ && $9671$ &  \\
& $[9,3,6]_{K}$ & $[c\bar{c}c\bar{b}]^{(\frac{1}{2})_{1}\oplus\frac{1}{2}}_{K_{2}}$ && $9685$ &  \\
& $[10,3,6]_{K}$ & $[c\bar{c}c\bar{b}]^{(\frac{1}{2})_{2}\oplus\frac{1}{2}}_{K_{2}}$ && $9673$ &  \\ 
& $[11,3,6]_{K}$ & $[c\bar{c}c\bar{b}]^{\frac{3}{2}\oplus\frac{1}{2}}_{K_{2}}$ && $9679$ &  \\
\hline \hline
&&~~~~~~~$IJ^{P}=02^{+}$&&\\ \hline
& $[i,j,k]_{\rm type}$ & Channel & $E_{th}$ & $E_{sc}$ & $E_{cc}$  \\ \hline
& $[6,1,1]_{m}$ & $[J/\psi B^{*}_{c}]^{1\oplus1}_{1 \bigotimes 1}$ & $9427$ & $9428$ & $ 9428 $ \\
& $[6,1,2]_{m}$ & $[J/\psi B^{*}_{c}]^{1\oplus1}_{8 \bigotimes 8}$ && $9664$ &  \\
& $[6,2,4]_{di}$ & $[cc\bar{c}\bar{b}]^{1\oplus1}_{\bar{3} \bigotimes 3}$ && $9701$ &  \\
& $[12,3,5]_{K}$ & $[c\bar{c}c\bar{b}]^{\frac{3}{2}\oplus\frac{1}{2}}_{K_{1}}$ && $9679$ &  \\
& $[12,3,6]_{K}$ & $[c\bar{c}c\bar{b}]^{\frac{3}{2}\oplus\frac{1}{2}}_{K_{2}}$ && $9687$ &  \\
\hline\hline
\end{tabular}
\end{table}

\begin{table}
\caption{\label{tab:table4}The energies (in MeV) of the meson-meson structure for tetraquarks $bb\bar{b}\bar{c}$.}
\begin{tabular}{ccccccccccc}
\hline \hline
&&~~~~~~~~~~~~$IJ^{P}=00^{+}$&&\\ \hline
& $[i,j,k]_{\rm type}$ & Channel & $E_{th}$ & $E_{sc}$ & $E_{cc}$  \\ \hline
& $[1,1,1]_{m}$ & $[\eta_{b}B_{c}]^{0\oplus0}_{1 \bigotimes 1}$ & $15729$ & $15730$ & $15730 $ \\
& $[2,1,1]_{m}$ & $[\Upsilon B^{*}_{c}]^{1\oplus1}_{1 \bigotimes 1}$ & $15835$ & $15835$ &  \\
& $[1,1,2]_{m}$ & $[\eta_{b}B_{c}]^{0\oplus0}_{8 \bigotimes 8}$ & $$ & $16070$ &  \\
& $[2,1,2]_{m}$ & $[\Upsilon B^{*}_{c}]^{1\oplus1}_{8 \bigotimes 8}$ & $$ & $16071$ &  \\
& $[1,2,3]_{di}$ & $[bb\bar{b}\bar{c}]^{0\oplus0}_{6 \bigotimes \bar{6}}$ & $$ & $16059$ &  \\ 
& $[2,2,4]_{di}$ & $[bb\bar{b}\bar{c}]^{1\oplus1}_{\bar{3} \bigotimes 3}$& $$ & $16089$ &  \\
& $[7,3,5]_{K}$ &$[b\bar{b}b\bar{c}]^{(\frac{1}{2})_{1}\oplus\frac{1}{2}}_{K_{1}}$ & $$ & $15982$ &$  $ &\\
& $[8,3,5]_{K}$ & $[b\bar{b}b\bar{c}]^{(\frac{1}{2})_{2}\oplus\frac{1}{2}}_{K_{1}}$ & $$ & $16019$ &\\
& $[7,3,6]_{K}$ & $[b\bar{b}b\bar{c}]^{(\frac{1}{2})_{1}\oplus\frac{1}{2}}_{K_{2}}$ & $$ & $16077$ &  \\
& $[8,3,6]_{K}$ & $[b\bar{b}b\bar{c}]^{(\frac{1}{2})_{2}\oplus\frac{1}{2}}_{K_{2}}$ & $$ & $16069$ &  \\ 
\hline \hline
&&~~~~~~~~~~~~$IJ^{P}=01^{+}$&&\\ \hline
& $[i,j,k]_{\rm type}$ & Channel & $E_{th}$ & $E_{sc}$ & $E_{cc}$  \\ \hline
& $[4,1,1]_{m}$ & $[\Upsilon B_{c}]^{1\oplus0}_{1 \bigotimes 1}$ &$15781$ & $15781$ & $15781$ \\
& $[3,1,1]_{m}$ & $[\eta_{b}B^{*}_{c}]^{0\oplus1}_{1 \bigotimes 1}$ & $15784$ & $15785$ &  \\
& $[5,1,1]_{m}$ & $[\Upsilon B^{*}_{c}]^{1\oplus1}_{1 \bigotimes 1}$ & $15835$ & $15835$ &  \\
& $[3,1,2]_{m}$ & $[\eta_{b}B^{*}_{c}]^{0\oplus1}_{8 \bigotimes 8}$ & $$ & $16065$ &  \\
& $[4,1,2]_{m}$ & $[\Upsilon B_{c}]^{1\oplus0}_{8 \bigotimes 8}$ & $$ & $16066$ &  \\ 
& $[5,1,2]_{m}$ & $[\Upsilon B^{*}_{c}]^{1\oplus1}_{8 \bigotimes 8}$ & $$ & $16065$ &  \\
& $[3,2,3]_{di}$ & $[bb\bar{b}\bar{c}]^{0\oplus1}_{6 \bigotimes \bar{6}}$ & $$ & $16056$ &  \\
& $[4,2,4]_{di}$ & $[bb\bar{b}\bar{c}]^{1\oplus0}_{\bar{3} \bigotimes 3}$ & $$ & $16081$ &  \\
& $[5,2,4]_{di}$ & $[bb\bar{b}\bar{c}]^{1\oplus1}_{\bar{3} \bigotimes 3}$ & $$ & $16092$ &  \\
& $[9,3,5]_{K}$ & $[b\bar{b}b\bar{c}]^{(\frac{1}{2})_{1}\oplus\frac{1}{2}}_{K_{1}}$ & $$ & $15991$ &$$&  \\
& $[10,3,5]_{K}$ & $[b\bar{b}b\bar{c}]^{(\frac{1}{2})_{2}\oplus\frac{1}{2}}_{K_{1}}$ & $$ & $16011$ &  \\ 
& $[11,3,5]_{K}$ & $[b\bar{b}b\bar{c}]^{\frac{3}{2}\oplus\frac{1}{2}}_{K_{1}}$ & $$ & $16040$ &  \\
& $[9,3,6]_{K}$ & $[b\bar{b}b\bar{c}]^{(\frac{1}{2})_{1}\oplus\frac{1}{2}}_{K_{2}}$ & $$ & $16071$ &  \\
& $[10,3,6]_{K}$ & $[b\bar{b}b\bar{c}]^{(\frac{1}{2})_{2}\oplus\frac{1}{2}}_{K_{2}}$ & $$ & $16078$ &  \\ 
& $[11,3,6]_{K}$ & $[b\bar{b}b\bar{c}]^{\frac{3}{2}\oplus\frac{1}{2}}_{K_{2}}$ & $$ & $16060$ &  \\
\hline \hline
&&~~~~~~~~~~~~$IJ^{P}=02^{+}$&&\\ \hline
& $[i,j,k]_{\rm type}$ & Channel & $E_{th}$ & $E_{sc}$ & $E_{cc}$  \\ \hline
& $[6,1,1]_{m}$ & $[\Upsilon B^{*}_{c}]^{1\oplus1}_{1 \bigotimes 1}$ & $15835$ & $15835$ & $15835$ \\
& $[6,1,2]_{m}$ & $[\Upsilon B^{*}_{c}]^{1\oplus1}_{8 \bigotimes 8}$ & $$ & $16059$ &  \\
& $[6,2,4]_{di}$ & $[bb\bar{b}\bar{c}]^{1\oplus1}_{\bar{3} \bigotimes 3}$ & $$ & $16098$ &  \\
& $[12,3,5]_{K}$ & $[b\bar{b}b\bar{c}]^{\frac{3}{2}\oplus\frac{1}{2}}_{K_{1}}$ & $$ & $16050$ & & \\
& $[12,3,6]_{K}$ & $[b\bar{b}b\bar{c}]^{\frac{3}{2}\oplus\frac{1}{2}}_{K_{2}}$ & $$ & $16071$ &  \\
\hline\hline
\end{tabular}
\end{table}

There are no bound states in Tables~\ref{tab:table3} and \ref{tab:table4}. However, some resonance state could exist in the higher energy region. 
The ground and excited states of ``compact state'', i.e., color configurations $\bar{3} \bigotimes 3$, $6 \bigotimes \bar{6}$ and $8 \bigotimes 8$, could be the candidate for resonance state. To assess the stability of these resonance states, we use the real-scaling method(RSM)~\cite{Simons:1981}. 
The RSM is a way to find the possible resonance state. In this method, the Gaussian size parameters $r_{n}$ between two color-singlet sub-clusters are scaled by multiplying a factor $\alpha$, i.e., $r\rightarrow\alpha r$. 
The resonance state might exist if the avoid-crossing  structure appears repeatedly with the increasing $\alpha$. The repeated avoid-crossing  structures are caused by the interaction between the interaction of scattering state with higher energy and resonance state at larger distances.
 Then, the decay widths could be calculated by the following formula~\cite{Simons:1981}. 
\begin{eqnarray}
\Gamma=4 |V(\alpha_{c})|\frac{\sqrt{|S_{r}| |S_{c}|}}{|S_{c}-S_{r}|}\,. 
\end{eqnarray}
Here $V(\alpha_{c})$ stands for half of the minimal energy difference between resonance state and scattering state, while $S_{r}$ and $S_{c}$ represent the slope of the resonance state and the scattering state, respectively.

The avoid-crossing structures could be caused by two ways: (i) the interaction between the scattering and resonance states with the increasing $\alpha$, and these avoid-crossing structures would be possible resonance states, (ii) the interaction between two scattering states with different decay rates, which means its dominated component of the structure are scattering states, so these avoid-crossing structures couldn't regard as resonance states. 

To determine a state is a true resonance or not is whether it have resonance mechanism. For present study, because there is no bound state of meson-meson structure, the resonance mechanism is whether the avoid-crossing structures correspond to an excited state of ``compact state''.

\begin{figure}
\centering
\includegraphics[scale=0.255]{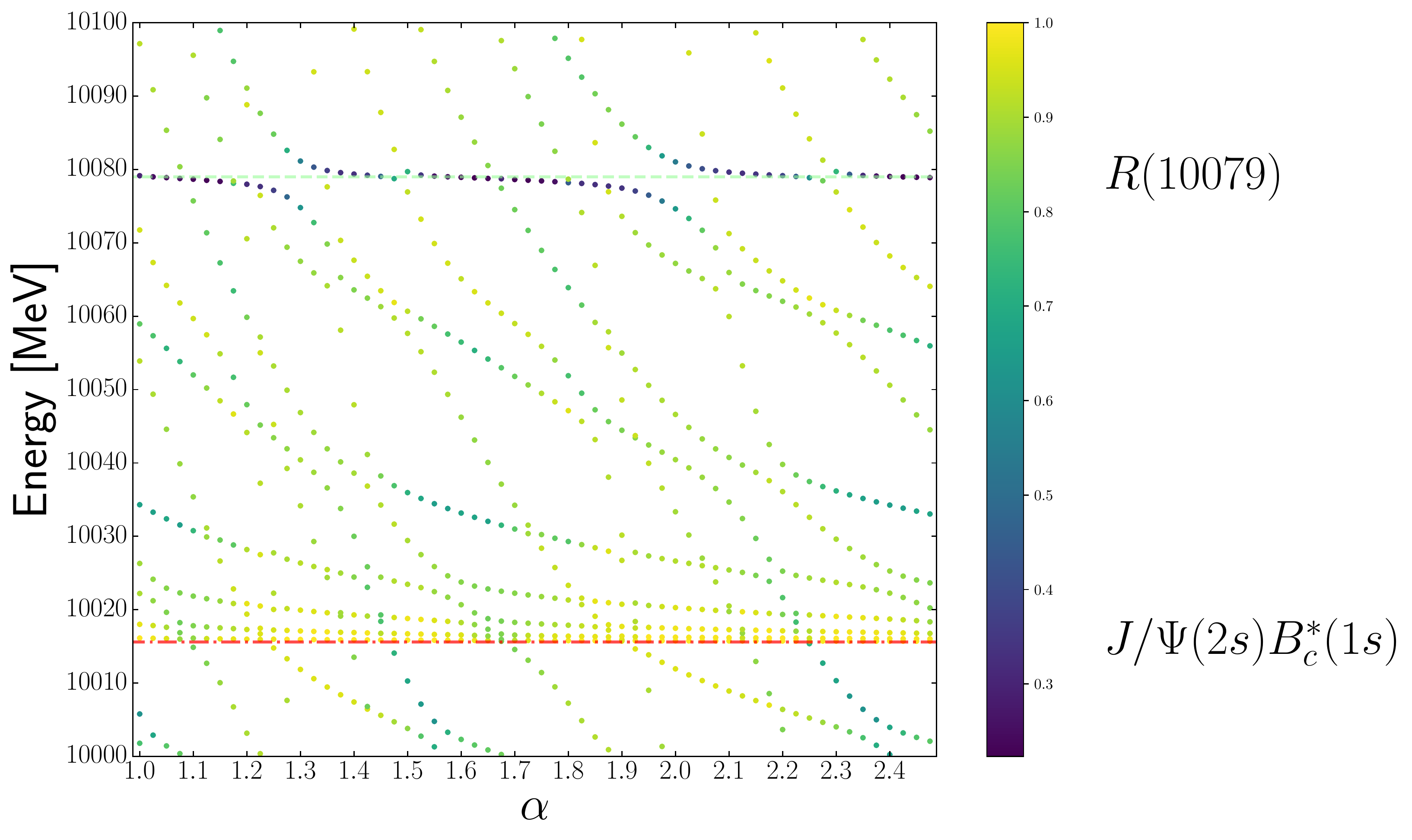}
\caption{Energy spectrum of $IJ^{P}=00^{+}$ in $cc\bar{c}\bar{b}$.}
\label{fig:cccb_00}
\end{figure}

\begin{figure}
\centering
\includegraphics[scale=0.255]{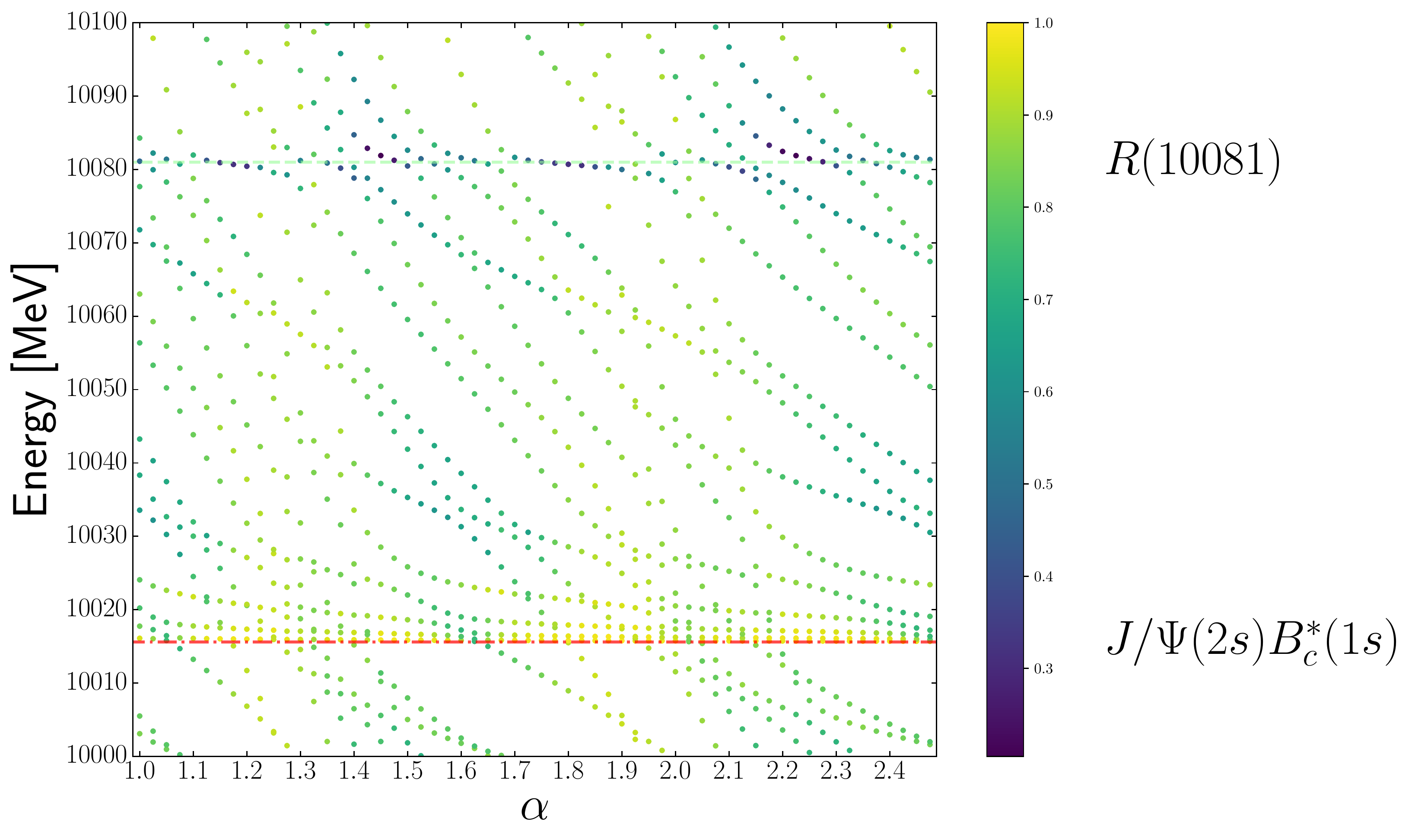}
\caption{Energy spectrum of $IJ^{P}=01^{+}$ in $cc\bar{c}\bar{b}$.}
\label{fig:cccb_01}
\end{figure}

\begin{figure}
\centering
\includegraphics[scale=0.255]{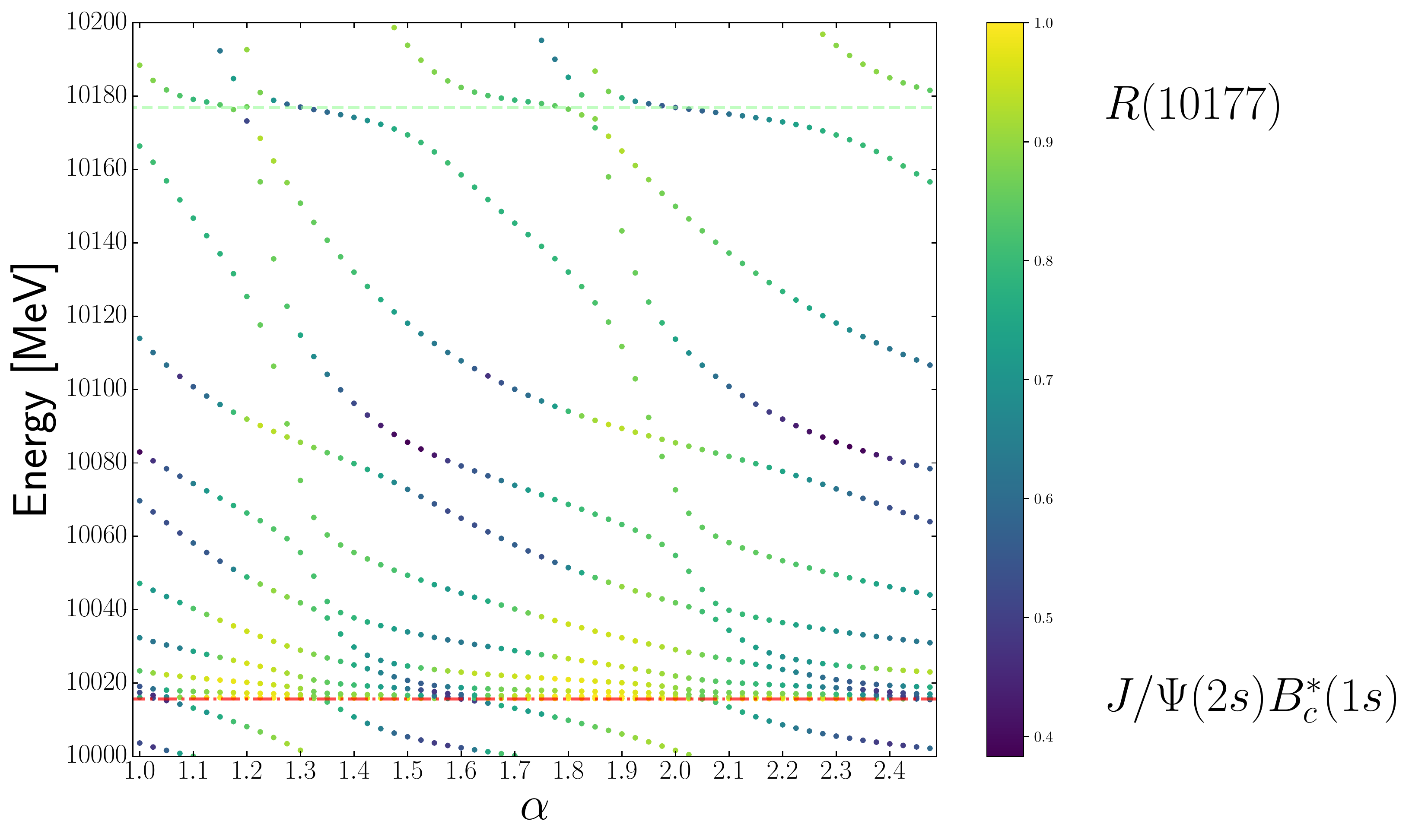}
\caption{Energy spectrum of $IJ^{P}=02^{+}$ in $cc\bar{c}\bar{b}$.}
\label{fig:cccb_02}
\end{figure}

\begin{figure}
	\centering
	\includegraphics[scale=0.255]{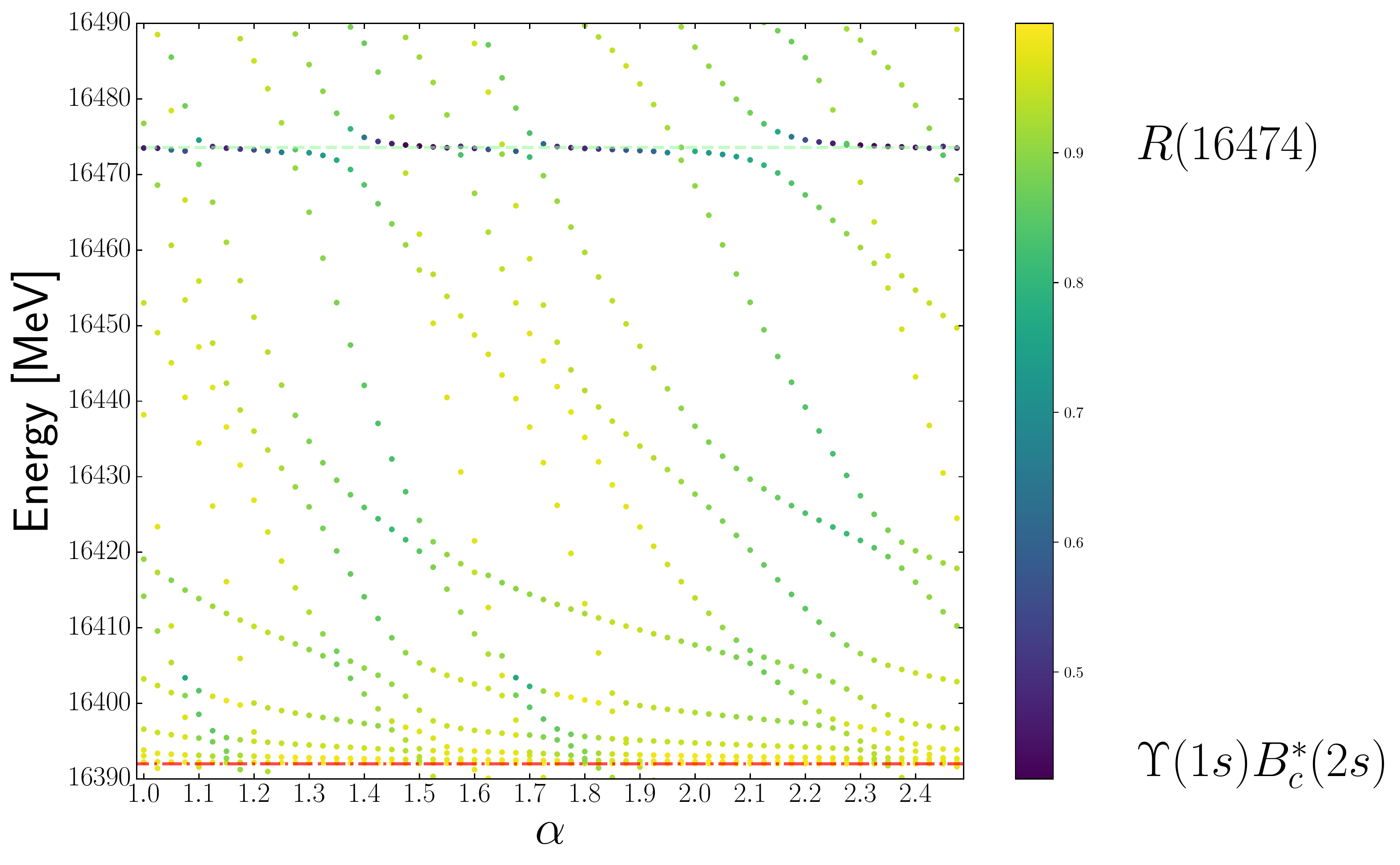}
	\caption{Energy spectrum of $IJ^{P}=00^{+}$ in $bb\bar{b}\bar{c}$.}
	\label{fig:bbbc_00}
\end{figure}

\begin{figure}
	\centering
	\includegraphics[scale=0.255]{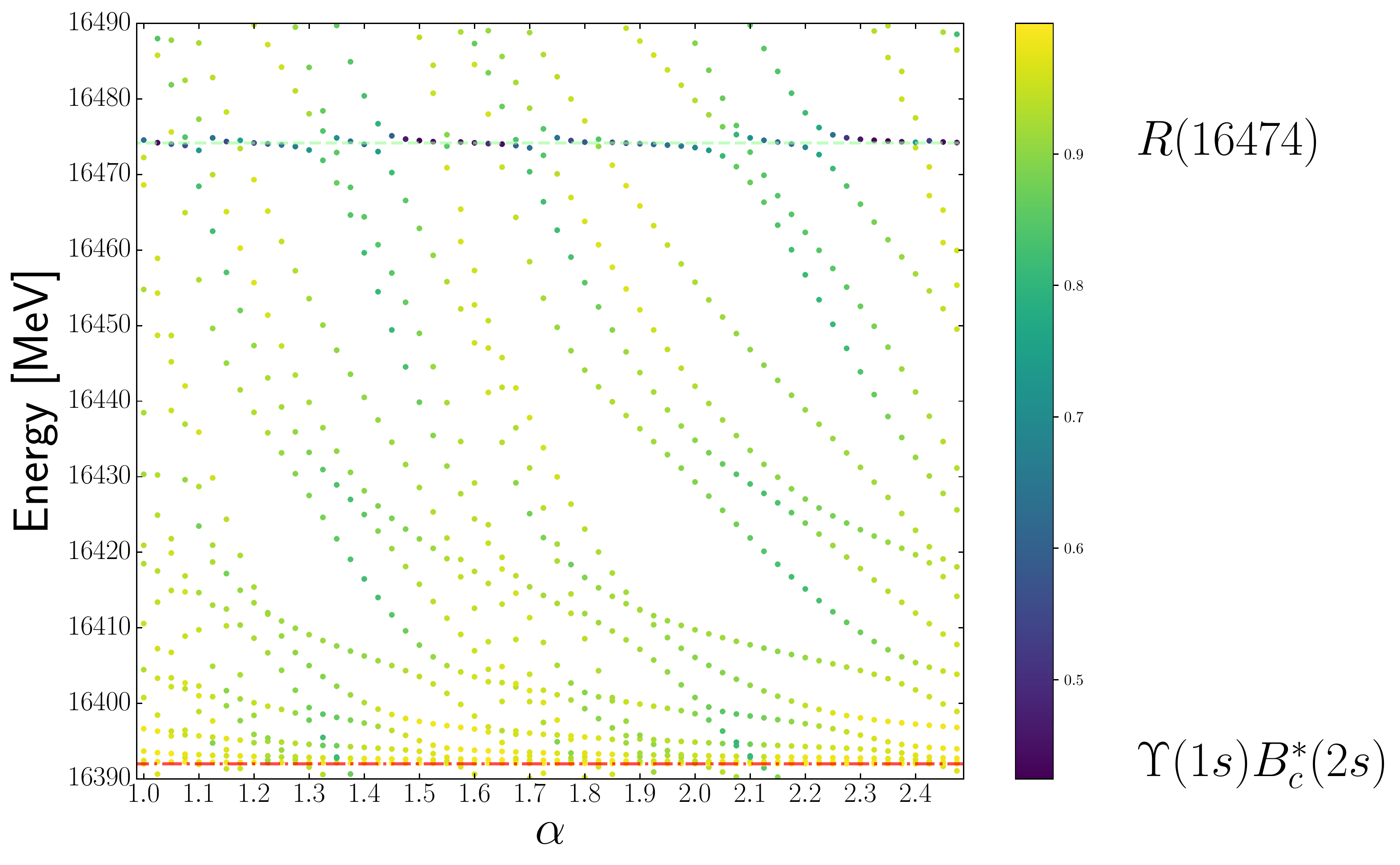}
	\caption{Energy spectrum of $IJ^{P}=01^{+}$ in $bb\bar{b}\bar{c}$.}
	\label{fig:bbbc_01}
\end{figure}

\begin{figure}
	\centering
	\includegraphics[scale=0.255]{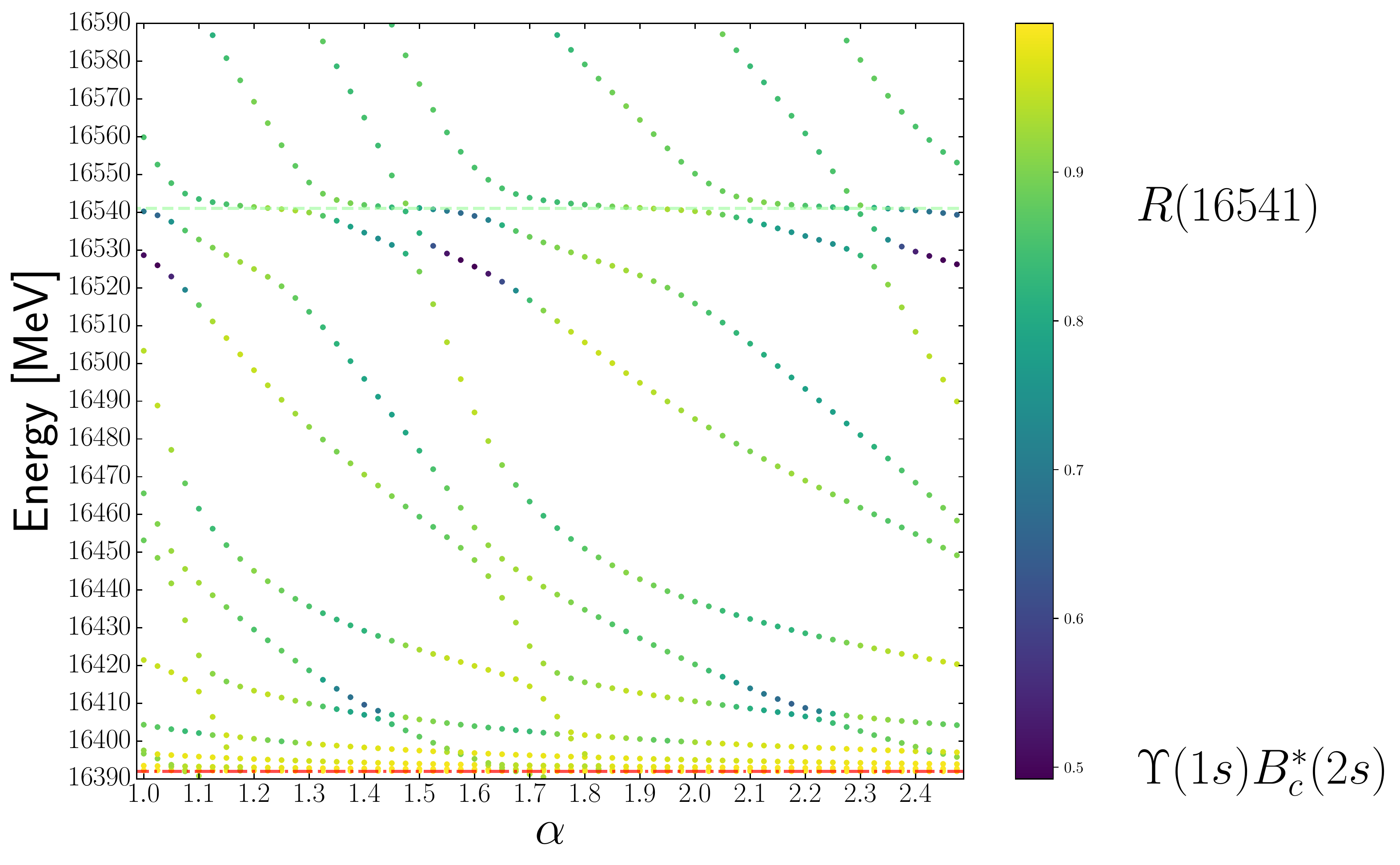}
	\caption{Energy spectrum of $IJ^{P}=02^{+}$ in $bb\bar{b}\bar{c}$.}
	\label{fig:bbbc_02}
\end{figure}

\begin{table}[!ht]
\centering
\caption{\label{tab:resonances}The max decay width of a single avoid-crossing structure $\Gamma({\rm MeV})$ and the total decay width $\Gamma_{total}({\rm MeV})$ of the resonance states in $cc\bar{c}\bar{b}$ and $bb\bar{b}\bar{c}$ systems. The last column means the energy level of the ``compact states''.}
\begin{tabular}{cccccc}
\hline\hline
System&$IJ^{P}$&Resonance state & $\Gamma$ & $\Gamma_{total}$& State \\\hline
\multirow{3}{*}{$cc\bar{c}\bar{b}$}&$00^{+}$&R(10079)&$ 6.7 $&$ 8.4 $&2S\\
&$01^{+}$&R(10081)&$ 1.4 $&$ 7.2 $&2S\\
&$02^{+}$&R(10177)&$ 9.1 $&$ 11.1 $&3S\\\hline	
\multirow{3}{*}{$bb\bar{b}\bar{c}$}&$00^{+}$&R(16474)&$ 2.2 $&$ 6.1 $&2S\\
&$01^{+}$&R(16474)&$ 2.2 $&$ 6.9 $&2S\\
&$02^{+}$&R(16541)&$ 5.3 $&$ 8.5 $&3S\\
\hline\hline
\end{tabular}
\end{table}

Figs.~\ref{fig:cccb_00}, \ref{fig:cccb_01} and \ref{fig:cccb_02} are the energy spectrum of $cc\bar{c}\bar{b}$ system, while Figs.~\ref{fig:bbbc_00}, \ref{fig:bbbc_01} and \ref{fig:bbbc_02} are the energy spectrum of $bb\bar{b}\bar{c}$ system. 
Every point shown in figures represent the energy calculated from channels coupling. 
The red dash-dote line is the threshold, and the corresponding physical channel is marked on the right of figure. 
The green dashed line refers to the mass of possible resonance state. 
The color of each point shows the value of $P_{1 \bigotimes 1}$.  
The legend on the right represents the value of $P_{1 \bigotimes 1}$, i.e., yellow and dark representing high and low percentage of $P_{1 \bigotimes 1}$. The percentage of $P_{8 \bigotimes 8}$ is obtained by $1-P_{1 \bigotimes 1}$.

From Figs.~\ref{fig:cccb_00}-\ref{fig:bbbc_02} we found that, lots of avoid-crossing structures appears in $IJ^{P}=00^{+}, 01^{+}$ and $02^{+}$ channels. 
However, only six states have resonance mechanism, i.e., they correspond to the excited states of their ``compact state''. 
We found 3 resonant states in $cc\bar{c}\bar{b}$ channel, i.e.,  $R(10079)$ in $IJ^{P}=00^{+}$, $R(10081)$ in $IJ^{P}=01^{+}$ and $R(10177)$ in $IJ^{P}=02^{+}$, and 3 resonant states in $bb\bar{b}\bar{c}$ channel, i.e.,  $R(16474)$ in $IJ^{P}=00^{+}$, $R(16474)$ in $IJ^{P}=01^{+}$ and $R(16541)$ in $IJ^{P}=02^{+}$. Each of these resonance states has more than 30\% of compact states. Table~\ref{tab:resonances} lists the decay width of the resonances states of $cc\bar{c}\bar{b}$ and $bb\bar{b}\bar{c}$ systems. 
There are many avoid-crossing structures on the resonant line, so we calculate the maximum decay width from a single avoid-crossing structure as the minimum total decay width of the system, and sum over all decay widths as a maximal total decay width. 

The 1S of ``compact states'' in $cc\bar{c}\bar{b}$ and $bb\bar{b}\bar{c}$ could not form a resonance through RSM. The reason would be that these 1S states are strongly coupled with scattering state and decay to threshold quickly. 
The 2S of ``compact states'' of $IJ^{P}=02^{+}$ in $cc\bar{c}\bar{b}$ and $bb\bar{b}\bar{c}$ systems are very close to the threshold of $J/\Psi(2S)B_c^*(1S)$ and $\Upsilon(1S)B_c^*(2S)$, so they couple to threshold very strong, and hard to form a resonance line.

%##########
\section{\label{sec:level4}summary}
We investigate the full-heavy tetraquarks $cc\bar{c}\bar{b}$ and $bb\bar{b}\bar{c}$ in meson-meson structure, diquark-antidiquark structure and K-structure within the framework of the quark model, and consider the combination of all possible color, flavor, spin configurations. In both of $cc\bar{c}\bar{b}$ and $bb\bar{b}\bar{c}$ systems, we couple all channels in quantum numbers of $ IJ^{P}=00^{+}$, $01^{+}$ and $ 02^{+}$, and found that there is no bound state exists. However, through RSM, we found three possible resonance states $R(10079)$, $R(10081)$ and $R(10177)$ in $cc\bar{c}\bar{b}$ system, and three possible resonance states $R(16474)$, $R(16474)$ and $R(16541)$ in $bb\bar{b}\bar{c}$ system. The decay width of $R(10079)$, $R(10081)$ and $R(10177)$ are $6.7-8.4~{\rm MeV}$, $1.4-7.2~{\rm MeV}$ and $9.1-11.1~{\rm MeV}$, while the decay width of $R(16474)$, $R(16474)$ and $R(16541)$ are $2.2-6.1~{\rm MeV}$, $2.2-6.9~{\rm MeV}$ and $5.3-8.5~{\rm MeV}$. 
$bc\bar{c}\bar{c}$ and $cb\bar{b}\bar{b}$ systems will have the same results as $cc\bar{c}\bar{b}$ and $bb\bar{b}\bar{c}$, respectively.

For the present study, the annihilation interaction is inversely proportional to the masses of the interacting quarks, and the one-gluon-annihilation process must be very weak for heavy quarks~\cite{Yang:2017prf}. Therefore, the rearrangement decay is the mainly decay behaviors if the energy below its reference thresholds of $ D^+ D^- B_{c}(10014\rm MeV) $, $ D^+ B^+ \eta_{c}(10131\rm MeV) $ for $cc\bar{c}\bar{b}$ system and $ B^+B^-B_{c}(16834\rm MeV) $, $ D^-B^-\eta_{b}(16548\rm MeV) $ for $bb\bar{b}\bar{c}$ system. 
In this means, $R(10079)$, $R(10081)$ and $R(10177)$ could be found in $ D^+ D^- B_{c}(10014\rm MeV) $ channel. 
The full consideration of 4-6 mixing could be calculated in the future works.

M.S Liu \textit{et al.}~\cite{Liu:2019zuc} and G. Yang \textit{et al.}~\cite{Yang:2021hrb} also studied similar system. 
To evaluate the parameter dependence of each model, we calculate the error $ m_{err} $ by:
\begin{eqnarray}
	m_{err}=\sqrt{\dfrac{\sum_{i}(\bar{m}-m_i)^2}{(N-1)}}.
\end{eqnarray}
Here $ m_i $ refers to the energy of meson with each parameter has changed $\pm1\%$, while $ \bar{m} $ represents the average energy value of each parameter(except $m_c $ and $ m_b$) changed by 1\%, and ``N'' means the number of changed parameters from three models. The value of ``N'' in present study, Ref.~\cite{Liu:2019zuc} and Ref.~\cite{Yang:2021hrb} is 14, 14 and 12, respectively.

\begin{table}[ht]
	\centering
\caption{\label{tab:para1}
Theoretical and experimental mass of twelve mesons of present work and Refs.~\cite{Liu:2019zuc,Yang:2021hrb}}
\scalebox{0.82}{
\begin{tabular}{cccccc}
	\hline\hline
	&This paper & Ref.~\cite{Liu:2019zuc} & Ref.~\cite{Yang:2021hrb}&Exp \\ \hline
	$ \eta_c(1s) $ & $ 2989.2\pm4.3  $ & $ 2983.4\pm2.7 $  & $ 2968.1\pm2.0 $&$ 2983.9\pm0.4 $  \\ \hline
	$ \eta_c(2s)  $& $ 3626.8\pm5.0 $  & $ 3634.5\pm2.2 $  & $ 3655.1\pm2.9 $&$ 3637.5\pm1.1 $ \\ \hline
	$ J/\psi(1s) $ & $ 3096.6\pm4.1 $  & $ 3097.1\pm2.0$  & $ 3102.8\pm1.8 $ &$3096.900\pm0.006$ \\ \hline
	$ \psi(2s) $ & $ 3685.0\pm5.1 $  & $ 3678.5\pm2.2 $  & $ 3721.5\pm3.0 $&$ 3686.10\pm0.06 $  \\ \hline
	$ B_c(1s) $ & $ 6276.2\pm4.0 $  & $ 6270.5\pm2.7 $  & $ 6275.3\pm1.8 $&$ 6274.47\pm0.32 $  \\ \hline
	$ B_c(2s) $ & $ 6856.9\pm4.8 $  & $ 6870.7\pm2.0 $  & $ 6883.3\pm2.5 $&$ 6871.2\pm1.0 $  \\ \hline
	$ \eta_b(1s) $ & $ 9453.6\pm4.0  $ & $ 9389.5\pm3.6  $ & $ 9401.4\pm2.2 $&$ 9398.7\pm2.0 $  \\ \hline
	$ \eta_b(2s) $ & $ 9985.4\pm4.4 $  & $ 10004.7\pm1.8 $  & $ 9961.1\pm2.0 $&$ 9999.0\pm4.3  $ \\ \hline
	$ \Upsilon(1s) $ & $ 9504.5\pm3.8 $  & $ 9459.4\pm3.0 $ &$  9463.1\pm2.1 $ & $ 9460.30\pm0.26 $  \\ \hline
	$ \Upsilon(2s) $ & $ 10012.6\pm4.4 $  & $ 10024.0\pm1.8 $  & $ 9981.7\pm2.0 $&$ 10023.26\pm0.31 $  \\ \hline
	$ \Upsilon(3s) $ & $ 10335.0\pm5.0 $  & $ 10335.8\pm2.0 $  & $ 10330.1\pm2.6 $&$ 10355.2\pm0.5 $  \\ \hline
	$ \Upsilon(4s) $ & $ 10557.3\pm5.4 $  & $ 10582.2\pm2.4 $ & $ 10618.3\pm3.2 $&$ 10579.4\pm1.2 $ \\\hline
	\hline
\end{tabular}}
\end{table}

Table~\ref{tab:para1} lists theoretical and experimental mass of twelve mesons of present work and Refs.~\cite{Liu:2019zuc,Yang:2021hrb}, we found that three model can explain experiment with a parameter error about $3-5$~MeV. 

The main difference of present work and Refs.~\cite{Liu:2019zuc,Yang:2021hrb} are:

\begin{enumerate}
\item M.S Liu \textit{et al.}~\cite{Liu:2019zuc} have studied the meson-meson structure and diquark-antidiquark structure in $ cc\bar{c}\bar{b} $ and $ bb\bar{b}\bar{c}\ $ systems by a quark model. However, there are two main differences between our study and theirs: (1) they use different potential of $V_{ij}^{CON}$ and $V_{ij}^{OGE}$, where 
$V_{ij}^{CON}(r_{ij})  = -\frac{3}{16}\boldsymbol{\lambda}^c_{i}\cdot
\boldsymbol{\lambda}^c_{j}\cdot br_{ij}$, and   
$V^{OGE}_{ij}(r_{ij})  =  \frac{\alpha_{ij}}{4}{\boldsymbol\lambda}^c_i \cdot
{\boldsymbol\lambda}^c_j
[\frac{1}{r_{ij}}-\frac{\pi}{2}\frac{\sigma^3_{ij}e^{-\sigma_{ij}^2r_{ij}^2}}{\pi^{3/2}}
\frac{4}{3m_{i}m_{j}}(
\boldsymbol{\sigma}_i\cdot\boldsymbol{\sigma}_j)]$.
Here $\alpha_{ij} $ and $ \sigma_{ij} $ are parameter related to the types of two quarks, and $ b $ is the strength of confinement. (2) Their results reveals that the 1S of diquark-antidiquark structure with sextet-antisextet and triplet-antitriplet from single channel calculation would be possible resonance state, but in present study 1S of diquark-antidiquark structure couldn't form a resonance through RSM. The reason would be that these 1S channels of diquark-antidiquark structure are strongly coupled with scattering state and decay to threshold quickly. 

\item G. Yang \textit{et al.}~\cite{Yang:2021hrb} employs a potential model inspired by the Lattice-QCD investigation of Ref.~\cite{Kawanai:2011jt} to study $ cc\bar{c}\bar{b} $ and $ bb\bar{b}\bar{c}\ $ systems, and the K-structure are also considered in their calculation. In their model, the gluonic potential is 
$ V_{ij}^{CON+OGE}(r_{ij}) =-\frac{3}{16}(\boldsymbol{\lambda}^c_{i}\cdot
\boldsymbol{\lambda}^c_{j})[ -\frac{\alpha}{r_{ij}}+\sigma r_{ij}+\beta e^{-\gamma r_{ij}} \frac{(
\boldsymbol{\sigma}_i\cdot\boldsymbol{\sigma}_j)}{4}]$, where the model parameters $ \alpha $, $ \beta $, $\gamma$, and $\sigma$ can be determined via a calculation of the mass spectrum of the S-wave $Q\bar{Q}$ mesons~\cite{Yang:2021hrb}, while the $V_{ij}^{CON}$ and $V_{ij}^{OGE}$ in our model are described as Eq.~(\ref{con}) and Eq.~(\ref{oge}). 
According to their results, they found 7 possible resonances in $ cc\bar{c}\bar{b} $ system and 3 possible resonances in $ bb\bar{b}\bar{c}\ $ system~\cite{Yang:2021hrb}.
According to the data in Table~\ref{tab:para1} and compared with their conclusions, the difference between our study and theirs mainly reflected in the following: (1) Firstly, they employ a linear confinement potential, while we use a confinement potential with screen effects, and the magnitude of attractive force comes from linear confinement are more than ours in medium and long range. Because the size of excited meson is larger than ground state meson, there model have more deviation between theoretical prediction and experiment result in excited state. 
(2) In addition, their conclusions points that all the solutions are possible resonance states, but does not analyze the resonance mechanism. We found that the excited states of ``compact states'' could be candidate of resonance state, and found the corresponding resonant line by RSM.
\end{enumerate}

Since current experimental data are not sufficient to determine which form of confinement potential is better, we expected that more experimental observation data in the future could help. 

\begin{table}[!ht]
\centering
\caption{\label{tab:K_M_excep}The energy $ E $ of R(10081) and the error $ E(1\%) $ calculated by change $\pm1\%$ of each parameter(except $m_c$ and $m_b$). }
\begin{tabular}{cccccc}
\hline\hline
Resonance state & $ E $ & $ E(1\%) $& \\\hline
R(10081)&$ 10080.5 $&$ 10079.5\pm10.7 $&\\
\hline\hline
\end{tabular}
\end{table}

To evaluate the stability of possible resonance states, we calculate the energy $ E $ of R(10081) and the error $ E(1\%) $ calculated by change $\pm1\%$ of each parameter(except $m_c$ and $m_b$), which is list in Table~\ref{tab:K_M_excep}.  
According to Tables~\ref{tab:para1} and~\ref{tab:K_M_excep} we conclude that the result of present model has little effects on the changing of parameters. 
 
In addition, all six possible resonance states are below $P-$wave meson composed of $c\bar{c}$ and $b\bar{b}$. In higher energy region, $P-$wave meson might play a significant role, which demands careful consideration. We leave it in the future works.

\begin{acknowledgments}
B.R.~He was supported in part by the National Natural Science Foundation of China (Grant Nos. 11705094 and 12047503), Natural Science Foundation of Jiangsu Province, China (Grant No. BK20171027), Natural Science Foundation of the Higher Education Institutions of Jiangsu Province, China (Grant Nos. 17KJB140011 and 22KJB140012).
And the work of J.L.~Ping was supported in part by the National Natural Science Foundation of China under Grants No. 11775118, and No. 11535005.
\end{acknowledgments}


\clearpage
\begin{thebibliography}{99}

%\cite{Gell-Mann:1964ewy}
\bibitem{Gell-Mann:1964ewy}
M.~Gell-Mann,
%``A Schematic Model of Baryons and Mesons,''
Phys. Lett. \textbf{8}, 214-215 (1964).
%doi:10.1016/S0031-9163(64)92001-3
%3628 citations counted in INSPIRE as of 08 Feb 2022

%\cite{Zweig:1964jf}
\bibitem{Zweig:1964jf}
G.~Zweig,
%``An SU(3) model for strong interaction symmetry and its breaking. Version 2,''
CERN-TH-412.
%585 citations counted in INSPIRE as of 08 Feb 2022

%\cite{Belle:2003nnu}
\bibitem{Belle:2003nnu}
S.~K.~Choi \textit{et al.} [Belle],
%``Observation of a narrow charmonium-like state in exclusive $B^\pm \to K^\pm \pi^+ \pi^- J/\psi$ decays,''
Phys. Rev. Lett. \textbf{91}, 262001 (2003).
%doi:10.1103/PhysRevLett.91.262001
%[arXiv:hep-ex/0309032 [hep-ex]].
%2071 citations counted in INSPIRE as of 08 Feb 2022

%\cite{CDF:2003cab}
\bibitem{CDF:2003cab}
D.~Acosta \textit{et al.} [CDF],
%``Observation of the narrow state $X(3872) \to J/\psi \pi^+ \pi^-$ in $\bar{p}p$ collisions at $\sqrt{s} = 1.96$ TeV,''
Phys. Rev. Lett. \textbf{93}, 072001 (2004).
%doi:10.1103/PhysRevLett.93.072001
%[arXiv:hep-ex/0312021 [hep-ex]].
%918 citations counted in INSPIRE as of 02 Dec 2022

%\cite{D0:2004zmu}
\bibitem{D0:2004zmu}
V.~M.~Abazov \textit{et al.} [D0],
%``Observation and properties of the $X(3872)$ decaying to $J/\psi \pi^+ \pi^-$ in $p\bar{p}$ collisions at $\sqrt{s} = 1.96$ TeV,''
Phys. Rev. Lett. \textbf{93}, 162002 (2004).
%doi:10.1103/PhysRevLett.93.162002
%[arXiv:hep-ex/0405004 [hep-ex]].
%800 citations counted in INSPIRE as of 02 Dec 2022

%\cite{BaBar:2004oro}
\bibitem{BaBar:2004oro}
B.~Aubert \textit{et al.} [BaBar],
%``Study of the $B \to J/\psi K^- \pi^+ \pi^-$ decay and measurement of the $B \to X(3872) K^-$ branching fraction,''
Phys. Rev. D \textbf{71}, 071103 (2005). 
%doi:10.1103/PhysRevD.71.071103
%[arXiv:hep-ex/0406022 [hep-ex]].
%722 citations counted in INSPIRE as of 02 Dec 2022

%%\cite{QuarkoniumWorkingGroup:2004kpm}
%\bibitem{QuarkoniumWorkingGroup:2004kpm}
%N.~Brambilla \textit{et al.} [Quarkonium Working Group],
%%``Heavy quarkonium physics,''

%\cite{Brambilla:2010cs}
\bibitem{Brambilla:2010cs}
N.~Brambilla, S.~Eidelman, B.~K.~Heltsley, R.~Vogt, G.~T.~Bodwin, E.~Eichten, A.~D.~Frawley, A.~B.~Meyer, R.~E.~Mitchell and V.~Papadimitriou, \textit{et al.}
%``Heavy Quarkonium: Progress, Puzzles, and Opportunities,''
Eur. Phys. J. C \textbf{71}, 1534 (2011). 
%doi:10.1140/epjc/s10052-010-1534-9
%[arXiv:1010.5827 [hep-ph]].
%1736 citations counted in INSPIRE as of 02 Dec 2022

%\cite{Faccini:2012pj}
\bibitem{Faccini:2012pj}
R.~Faccini, A.~Pilloni and A.~D.~Polosa,
%``Exotic Heavy Quarkonium Spectroscopy: A Mini-review,''
Mod. Phys. Lett. A \textbf{27}, 1230025 (2012). 
%doi:10.1142/S021773231230025X
%[arXiv:1209.0107 [hep-ph]].
%53 citations counted in INSPIRE as of 02 Dec 2022

%\cite{Chen:2016qju}
\bibitem{Chen:2016qju}
H.~X.~Chen, W.~Chen, X.~Liu and S.~L.~Zhu,
%``The hidden-charm pentaquark and tetraquark states,''
Phys. Rept. \textbf{639}, 1-121 (2016). 
%doi:10.1016/j.physrep.2016.05.004
%[arXiv:1601.02092 [hep-ph]].
%890 citations counted in INSPIRE as of 02 Dec 2022

%\cite{Esposito:2014rxa}
\bibitem{Esposito:2014rxa}
A.~Esposito, A.~L.~Guerrieri, F.~Piccinini, A.~Pilloni and A.~D.~Polosa,
%``Four-Quark Hadrons: an Updated Review,''
Int. J. Mod. Phys. A \textbf{30}, 1530002 (2015). 
%doi:10.1142/S0217751X15300021
%[arXiv:1411.5997 [hep-ph]].
%221 citations counted in INSPIRE as of 02 Dec 2022 

%\cite{Esposito:2016noz}
\bibitem{Esposito:2016noz}
A.~Esposito, A.~Pilloni and A.~D.~Polosa,
%``Multiquark Resonances,''
Phys. Rept. \textbf{668}, 1-97 (2017). 
%doi:10.1016/j.physrep.2016.11.002
%[arXiv:1611.07920 [hep-ph]].
%521 citations counted in INSPIRE as of 02 Dec 2022

%\cite{Chen:2016spr}
\bibitem{Chen:2016spr}
H.~X.~Chen, W.~Chen, X.~Liu, Y.~R.~Liu and S.~L.~Zhu,
%``A review of the open charm and open bottom systems,''
Rept. Prog. Phys. \textbf{80}, no.7, 076201 (2017). 
%doi:10.1088/1361-6633/aa6420
%[arXiv:1609.08928 [hep-ph]].
%288 citations counted in INSPIRE as of 02 Dec 2022

%\cite{Guo:2017jvc}
\bibitem{Guo:2017jvc}
F.~K.~Guo, C.~Hanhart, U.~G.~Mei\ss{}ner, Q.~Wang, Q.~Zhao and B.~S.~Zou,
%``Hadronic molecules,''
Rev. Mod. Phys. \textbf{90}, no.1, 015004 (2018). 
[erratum: Rev. Mod. Phys. \textbf{94}, no.2, 029901 (2022)].
%doi:10.1103/RevModPhys.90.015004
%[arXiv:1705.00141 [hep-ph]].
%844 citations counted in INSPIRE as of 02 Dec 2022


%\cite{Olsen:2017bmm}
\bibitem{Olsen:2017bmm}
S.~L.~Olsen, T.~Skwarnicki and D.~Zieminska,
%``Nonstandard heavy mesons and baryons: Experimental evidence,''
Rev. Mod. Phys. \textbf{90}, no.1, 015003 (2018).
%doi:10.1103/RevModPhys.90.015003
%[arXiv:1708.04012 [hep-ph]].
%527 citations counted in INSPIRE as of 03 Dec 2022


%\cite{Liu:2019zoy}
\bibitem{Liu:2019zoy}
Y.~R.~Liu, H.~X.~Chen, W.~Chen, X.~Liu and S.~L.~Zhu,
%``Pentaquark and Tetraquark states,''
Prog. Part. Nucl. Phys. \textbf{107}, 237-320 (2019).
%doi:10.1016/j.ppnp.2019.04.003
%[arXiv:1903.11976 [hep-ph]].
%416 citations counted in INSPIRE as of 03 Dec 2022


%\cite{Brambilla:2019esw}
\bibitem{Brambilla:2019esw}
N.~Brambilla, S.~Eidelman, C.~Hanhart, A.~Nefediev, C.~P.~Shen, C.~E.~Thomas, A.~Vairo and C.~Z.~Yuan,
%``The $XYZ$ states: experimental and theoretical status and perspectives,''
Phys. Rept. \textbf{873}, 1-154 (2020).
%doi:10.1016/j.physrep.2020.05.001
%[arXiv:1907.07583 [hep-ex]].
%454 citations counted in INSPIRE as of 03 Dec 2022

%\cite{Chen:2022asf}
\bibitem{Chen:2022asf}
H.~X.~Chen, W.~Chen, X.~Liu, Y.~R.~Liu and S.~L.~Zhu,
%``An updated review of the new hadron states,''
Rept. Prog. Phys. \textbf{86}, no.2, 026201 (2023).
%doi:10.1088/1361-6633/aca3b6
%[arXiv:2204.02649 [hep-ph]].
%147 citations counted in INSPIRE as of 30 May 2023




%\cite{CMS:2016liw}
\bibitem{CMS:2016liw}
V.~Khachatryan \textit{et al.} [CMS],
%``Observation of $\Upsilon$(1S) pair production in proton-proton collisions at $ \sqrt{s}=8 $ TeV,''
JHEP \textbf{05}, 013 (2017).
%doi:10.1007/JHEP05(2017)013
%[arXiv:1610.07095 [hep-ex]].
%101 citations counted in INSPIRE as of 03 Dec 2022



%\cite{LHCb:2018uwm}
\bibitem{LHCb:2018uwm}
R.~Aaij \textit{et al.} [LHCb],
%``Search for beautiful tetraquarks in the $\Upsilon(1S)\mu^+\mu^-$ invariant-mass spectrum,''
JHEP \textbf{10}, 086 (2018).
%doi:10.1007/JHEP10(2018)086
%[arXiv:1806.09707 [hep-ex]].
%68 citations counted in INSPIRE as of 03 Dec 2022


%\cite{LHCb:2011kri}
\bibitem{LHCb:2011kri}
R.~Aaij \textit{et al.} [LHCb],
%``Observation of $J/\psi$ pair production in $pp$ collisions at $\sqrt{s}=7 TeV$,''
Phys. Lett. B \textbf{707}, 52-59 (2012).
%doi:10.1016/j.physletb.2011.12.015
%[arXiv:1109.0963 [hep-ex]].
%212 citations counted in INSPIRE as of 08 Feb 2022


%\cite{Belle:2002tfa}
\bibitem{Belle:2002tfa}
K.~Abe \textit{et al.} [Belle],
%``Observation of double c anti-c production in e+ e- annihilation at s**(1/2) approximately 10.6-GeV,''
Phys. Rev. Lett. \textbf{89}, 142001 (2002).
%doi:10.1103/PhysRevLett.89.142001
%[arXiv:hep-ex/0205104 [hep-ex]].
%373 citations counted in INSPIRE as of 08 Feb 2022

%\cite{LHCb:2020bwg}
\bibitem{LHCb:2020bwg}
R.~Aaij \textit{et al.} [LHCb],
%``Observation of structure in the $J /\psi$ -pair mass spectrum,''
Sci. Bull. \textbf{65}, no.23, 1983-1993 (2020).
%doi:10.1016/j.scib.2020.08.032
%[arXiv:2006.16957 [hep-ex]].
%231 citations counted in INSPIRE as of 03 Dec 2022

%\cite{CMS:2022yhl}
\bibitem{CMS:2022yhl}
[CMS],
%``Observation of new structures in the $\mathrm{J}/\psi \mathrm{J}/\psi$  mass spectrum in $\mathrm{p}\mathrm{p}$ collisions at $\sqrt{s} = 13$\textbackslash{},TeV,''
CMS-PAS-BPH-21-003.
%9 citations counted in INSPIRE as of 08 May 2023

%\cite{Iwasaki:1975pv}
\bibitem{Iwasaki:1975pv}
Y.~Iwasaki,
%``A Possible Model for New Resonances-Exotics and Hidden Charm,''
Prog. Theor. Phys. \textbf{54}, 492 (1975).
%doi:10.1143/PTP.54.492
%63 citations counted in INSPIRE as of 08 Feb 2022


%\cite{Lloyd:2003yc}
\bibitem{Lloyd:2003yc}
R.~J.~Lloyd and J.~P.~Vary,
%``All charm tetraquarks,''
Phys. Rev. D \textbf{70}, 014009 (2004).
%doi:10.1103/PhysRevD.70.014009
%[arXiv:hep-ph/0311179 [hep-ph]].
%81 citations counted in INSPIRE as of 03 Dec 2022



%\cite{Berezhnoy:2011xn}
\bibitem{Berezhnoy:2011xn}
A.~V.~Berezhnoy, A.~V.~Luchinsky and A.~A.~Novoselov,
%``Tetraquarks Composed of 4 Heavy Quarks,''
Phys. Rev. D \textbf{86}, 034004 (2012).
%doi:10.1103/PhysRevD.86.034004
%[arXiv:1111.1867 [hep-ph]].
%106 citations counted in INSPIRE as of 03 Dec 2022


%\cite{Karliner:2016zzc}
\bibitem{Karliner:2016zzc}
M.~Karliner, S.~Nussinov and J.~L.~Rosner,
%``$Q Q \bar Q \bar Q$ states: masses, production, and decays,''
Phys. Rev. D \textbf{95}, no.3, 034011 (2017).
%doi:10.1103/PhysRevD.95.034011
%[arXiv:1611.00348 [hep-ph]].
%140 citations counted in INSPIRE as of 03 Dec 2022



%\cite{Chen:2016jxd}
\bibitem{Chen:2016jxd}
W.~Chen, H.~X.~Chen, X.~Liu, T.~G.~Steele and S.~L.~Zhu,
%``Hunting for exotic doubly hidden-charm/bottom tetraquark states,''
Phys. Lett. B \textbf{773}, 247-251 (2017).
%doi:10.1016/j.physletb.2017.08.034
%[arXiv:1605.01647 [hep-ph]].
%100 citations counted in INSPIRE as of 08 Feb 2022



%\cite{Wu:2016vtq}
\bibitem{Wu:2016vtq}
J.~Wu, Y.~R.~Liu, K.~Chen, X.~Liu and S.~L.~Zhu,
%``Heavy-flavored tetraquark states with the $QQ\bar{Q}\bar{Q}$ configuration,''
Phys. Rev. D \textbf{97}, no.9, 094015 (2018).
%doi:10.1103/PhysRevD.97.094015
%[arXiv:1605.01134 [hep-ph]].
%121 citations counted in INSPIRE as of 03 Dec 2022





%\cite{Anwar:2017toa}
\bibitem{Anwar:2017toa}
M.~N.~Anwar, J.~Ferretti, F.~K.~Guo, E.~Santopinto and B.~S.~Zou,
%``Spectroscopy and decays of the fully-heavy tetraquarks,''
Eur. Phys. J. C \textbf{78}, no.8, 647 (2018).
%doi:10.1140/epjc/s10052-018-6073-9
%[arXiv:1710.02540 [hep-ph]].
%111 citations counted in INSPIRE as of 03 Dec 2022



%\cite{Esposito:2018cwh}
\bibitem{Esposito:2018cwh}
A.~Esposito and A.~D.~Polosa,
%``A $bb\bar b\bar b$di-bottomonium at the LHC?,''
Eur. Phys. J. C \textbf{78}, no.9, 782 (2018).
%doi:10.1140/epjc/s10052-018-6269-z
%[arXiv:1807.06040 [hep-ph]].
%68 citations counted in INSPIRE as of 08 Feb 2022

%\cite{Wang:2019rdo}
\bibitem{Wang:2019rdo}
G.~J.~Wang, L.~Meng and S.~L.~Zhu,
%``Spectrum of the fully-heavy tetraquark state $QQ\bar Q' \bar Q'$,''
Phys. Rev. D \textbf{100}, no.9, 096013 (2019).
%doi:10.1103/PhysRevD.100.096013
%[arXiv:1907.05177 [hep-ph]].
%74 citations counted in INSPIRE as of 03 Dec 2022


%\cite{Liu:2019zuc}
\bibitem{Liu:2019zuc}
M.~S.~Liu, Q.~F.~L\"u, X.~H.~Zhong and Q.~Zhao,
%``All-heavy tetraquarks,''
Phys. Rev. D \textbf{100}, no.1, 016006 (2019).
%doi:10.1103/PhysRevD.100.016006
%[arXiv:1901.02564 [hep-ph]].
%93 citations counted in INSPIRE as of 03 Dec 2022



%\cite{Jin:2020jfc}
\bibitem{Jin:2020jfc}
X.~Jin, Y.~Xue, H.~Huang and J.~Ping,
%``Full-heavy tetraquarks in constituent quark models,''
Eur. Phys. J. C \textbf{80}, no.11, 1083 (2020).
%doi:10.1140/epjc/s10052-020-08650-z
%[arXiv:2006.13745 [hep-ph]].
%44 citations counted in INSPIRE as of 08 Feb 2022

%\cite{Wang:2017jtz}
\bibitem{Wang:2017jtz}
Z.~G.~Wang,
%``Analysis of the $QQ\bar{Q}\bar{Q}$ tetraquark states with QCD sum rules,''
Eur. Phys. J. C \textbf{77}, no.7, 432 (2017).
%doi:10.1140/epjc/s10052-017-4997-0
%[arXiv:1701.04285 [hep-ph]].
%104 citations counted in INSPIRE as of 03 Dec 2022


%\cite{Yang:2020atz}
\bibitem{Yang:2020atz}
G.~Yang, J.~Ping and J.~Segovia,
%``Tetra- and penta-quark structures in the constituent quark model,''
Symmetry \textbf{12}, no.11, 1869 (2020).
%doi:10.3390/sym12111869
%[arXiv:2009.00238 [hep-ph]].
%54 citations counted in INSPIRE as of 03 Dec 2022

%\cite{Lu:2020cns}
\bibitem{Lu:2020cns}
Q.~F.~L\"u, D.~Y.~Chen and Y.~B.~Dong,
%``Masses of fully heavy tetraquarks $QQ {\bar{Q}} {\bar{Q}}$ in an extended relativized quark model,''
Eur. Phys. J. C \textbf{80}, no.9, 871 (2020).
%doi:10.1140/epjc/s10052-020-08454-1
%[arXiv:2006.14445 [hep-ph]].
%77 citations counted in INSPIRE as of 03 Dec 2022 


%\cite{Wan:2020fsk}
\bibitem{Wan:2020fsk}
B.~D.~Wan and C.~F.~Qiao,
%``Gluonic tetracharm configuration of $X (6900)$,''
Phys. Lett. B \textbf{817}, 136339 (2021).
%doi:10.1016/j.physletb.2021.136339
%[arXiv:2012.00454 [hep-ph]].
%38 citations counted in INSPIRE as of 03 Dec 2022

%\cite{Yang:2021hrb}
\bibitem{Yang:2021hrb}
G.~Yang, J.~Ping and J.~Segovia,
%``Exotic resonances of fully-heavy tetraquarks in a lattice-QCD insipired quark model,''
Phys. Rev. D \textbf{104}, no.1, 014006 (2021).
%doi:10.1103/PhysRevD.104.014006
%[arXiv:2104.08814 [hep-ph]].
%20 citations counted in INSPIRE as of 03 Dec 2022

%\cite{Wu:2022qwd}
\bibitem{Wu:2022qwd}
R.~H.~Wu, Y.~S.~Zuo, C.~Y.~Wang, C.~Meng, Y.~Q.~Ma and K.~T.~Chao,
%``NLO results with operator mixing for fully heavy tetraquarks in QCD sum rules,''
JHEP \textbf{11}, 023 (2022).
%doi:10.1007/JHEP11(2022)023
%[arXiv:2201.11714 [hep-ph]].
%9 citations counted in INSPIRE as of 03 Dec 2022 


%\cite{Zhao:2020zjh}
\bibitem{Zhao:2020zjh}
Z.~Zhao, K.~Xu, A.~Kaewsnod, X.~Liu, A.~Limphirat and Y.~Yan,
%``Study of charmoniumlike and fully-charm tetraquark spectroscopy,''
Phys. Rev. D \textbf{103}, no.11, 116027 (2021).
%doi:10.1103/PhysRevD.103.116027
%[arXiv:2012.15554 [hep-ph]].
%32 citations counted in INSPIRE as of 03 Dec 2022

%\cite{Wang:2021kfv}
\bibitem{Wang:2021kfv}
G.~J.~Wang, L.~Meng, M.~Oka and S.~L.~Zhu,
%``Higher fully charmed tetraquarks: Radial excitations and P-wave states,''
Phys. Rev. D \textbf{104}, no.3, 036016 (2021).
%doi:10.1103/PhysRevD.104.036016
%[arXiv:2105.13109 [hep-ph]].
%18 citations counted in INSPIRE as of 03 Dec 2022



%\cite{Liu:2021rtn}
\bibitem{Liu:2021rtn}
F.~X.~Liu, M.~S.~Liu, X.~H.~Zhong and Q.~Zhao,
%``Higher mass spectra of the fully-charmed and fully-bottom tetraquarks,''
Phys. Rev. D \textbf{104}, no.11, 116029 (2021).
%doi:10.1103/PhysRevD.104.116029
%[arXiv:2110.09052 [hep-ph]].
%14 citations counted in INSPIRE as of 03 Dec 2022


%\cite{Asadi:2021ids}
\bibitem{Asadi:2021ids}
Z.~Asadi and G.~R.~Boroun,
%``Masses of fully heavy tetraquark states from a four-quark static potential model,''
Phys. Rev. D \textbf{105}, no.1, 014006 (2022).
%doi:10.1103/PhysRevD.105.014006
%[arXiv:2112.11028 [hep-ph]].
%8 citations counted in INSPIRE as of 03 Dec 2022

%%\cite{Wang:2022yes}
%\bibitem{Wang:2022yes}
%G.~J.~Wang, Q.~Meng and M.~Oka,
%%``S-wave fully charmed tetraquark resonant states,''
%Phys. Rev. D \textbf{106}, no.9, 096005 (2022)
%%doi:10.1103/PhysRevD.106.096005
%%[arXiv:2208.07292 [hep-ph]].
%%5 citations counted in INSPIRE as of 02 Dec 2022

%\cite{Godfrey:1985xj}
\bibitem{Godfrey:1985xj}
S.~Godfrey and N.~Isgur,
%``Mesons in a Relativized Quark Model with Chromodynamics,''
Phys. Rev. D \textbf{32}, 189-231 (1985).
%doi:10.1103/PhysRevD.32.189
%3052 citations counted in INSPIRE as of 03 Dec 2022

%\cite{Deng:2017xlb}
\bibitem{Deng:2017xlb}
C.~Deng, J.~Ping, H.~Huang and F.~Wang,
%``Hidden charmed states and multibody color flux-tube dynamics,''
Phys. Rev. D \textbf{98}, no.1, 014026 (2018).
%doi:10.1103/PhysRevD.98.014026
%[arXiv:1801.00164 [hep-ph]].
%18 citations counted in INSPIRE as of 03 Dec 2022




%\cite{Vijande:2004he}
\bibitem{Vijande:2004he}
J.~Vijande, F.~Fernandez and A.~Valcarce,
%``Constituent quark model study of the meson spectra,''
J. Phys. G \textbf{31}, 481 (2005).
%doi:10.1088/0954-3899/31/5/017
%[arXiv:hep-ph/0411299 [hep-ph]].
%320 citations counted in INSPIRE as of 03 Dec 2022

%\cite{ParticleDataGroup:2022pth}
\bibitem{ParticleDataGroup:2022pth}
R.~L.~Workman \textit{et al.} [Particle Data Group],
%``Review of Particle Physics,''
PTEP \textbf{2022}, 083C01 (2022).
%doi:10.1093/ptep/ptac097
%190 citations counted in INSPIRE as of 05 Nov 2022

%\cite{Hiyama:2003cu}
\bibitem{Hiyama:2003cu}
E.~Hiyama, Y.~Kino and M.~Kamimura,
%``Gaussian expansion method for few-body systems,''
Prog. Part. Nucl. Phys. \textbf{51}, 223-307 (2003).
%doi:10.1016/S0146-6410(03)90015-9
%505 citations counted in INSPIRE as of 03 Dec 2022

\bibitem{Simons:1981}
J. Simons, J. Chem. Phys. 75, 2465 (1981).

%\cite{Yang:2017prf}
\bibitem{Yang:2017prf}
Y.~C.~Yang, Z.~Y.~Tan, J.~Ping and H.~S.~Zong,
%``Possible $D^{(*)}\bar{D}^{(*)}$ and $B^{(*)}\bar{B}^{(*)}$ molecular states in the extended constituent quark models,''
Eur. Phys. J. C \textbf{77}, no.9, 575 (2017).
%doi:10.1140/epjc/s10052-017-5137-6
%[arXiv:1703.09718 [hep-ph]].
%12 citations counted in INSPIRE as of 03 Dec 2022

%\cite{Kawanai:2011jt}
\bibitem{Kawanai:2011jt}
T.~Kawanai and S.~Sasaki,
%``Charmonium potential from full lattice QCD,''
Phys. Rev. D \textbf{85}, 091503 (2012).
%doi:10.1103/PhysRevD.85.091503
%[arXiv:1110.0888 [hep-lat]].
%66 citations counted in INSPIRE as of 02 Dec 2022


\end{thebibliography}
\end{document}